\documentclass[amsmath,10pt,nofootinbib,superscriptaddress,twocolumn,bibliography]{revtex4-1}
\usepackage{siunitx} 
\usepackage{microtype} 
\usepackage{amsmath}
\usepackage{amsbsy}
\usepackage{amssymb}
\usepackage{graphicx}
\usepackage{color}
\usepackage{physics}
\usepackage{bm}
\usepackage[capitalize]{cleveref}
\usepackage[normalem]{ulem} 
\usepackage{xr}
\makeatletter

\newcommand*{\addFileDependency}[1]{
  \typeout{(#1)}
  \@addtofilelist{#1}
  \IfFileExists{#1}{}{\typeout{No file #1.}}
}
\makeatother

\usepackage{verbatim}
\usepackage[bottom]{footmisc}

\makeatletter
\newcommand{\quickwordcount}[1]{%
  \immediate\write18{texcount -1 -sum -merge #1.tex > #1-words}%
  \immediate\openin\somefile=#1-words%
  \read\somefile to \@@localdummy%
  \immediate\closein\somefile%
  \setcounter{wordcounter}{\@@localdummy}%
  \@@localdummy%
}
\makeatother

\usepackage{natbib}

\begin{document}

\title{ Learning via mechanosensitivity and activity in cytoskeletal networks}


\author{Deb S. Banerjee}
\affiliation{James Franck Institute, University of Chicago, Chicago, IL 60637}

\author{Martin J. Falk}
\affiliation{James Franck Institute, University of Chicago, Chicago, IL 60637}
\affiliation{Department of Physics, University of Chicago, Chicago, IL 60637}


\author{Margaret L. Gardel}
\affiliation{James Franck Institute, University of Chicago, Chicago, IL 60637}
\affiliation{Department of Physics, University of Chicago, Chicago, IL 60637}
\affiliation{Department of Molecular Genetics and Cell Biology, University of Chicago, Chicago, IL 60637, USA}
\affiliation{Pritzker School for Molecular Engineering, The University of Chicago, Chicago, IL 60637, USA}

\author{Aleksandra M. Walczak}
\affiliation{Laboratoire de physique de l’École normale supérieure, CNRS, Paris Sciences et Lettres University, Sorbonne Université, and Université Paris-Cité, Paris, 75005, France}
\affiliation{James Franck Institute, University of Chicago, Chicago, IL 60637}
\affiliation{Department of Physics, University of Chicago, Chicago, IL 60637}

\author{Thierry Mora}
\affiliation{Laboratoire de physique de l’École normale supérieure, CNRS, Paris Sciences et Lettres University, Sorbonne Université, and Université Paris-Cité, Paris, 75005, France}
\affiliation{James Franck Institute, University of Chicago, Chicago, IL 60637}
\affiliation{Department of Physics, University of Chicago, Chicago, IL 60637}

\author{Suriyanarayanan Vaikuntanathan$^\dagger$}
\affiliation{James Franck Institute, University of Chicago, Chicago, IL 60637}
\affiliation{Department of Chemistry, University of Chicago, Chicago, IL 60637}


\begin{abstract}
In this work we show how a network inspired by a coarse-grained description of actomyosin cytoskeleton can learn - in a contrastive learning framework - from environmental perturbations if it is endowed with mechanosensitive proteins and motors. Our work is a proof of principle for how force-sensitive proteins and molecular motors can form the basis of a general strategy to learn in biological systems. Our work identifies a minimal biologically plausible learning mechanism and also explores its implications for commonly occuring phenomenolgy such as adaptation and homeostatis. 



\end{abstract}


\maketitle

\begingroup\renewcommand\thefootnote{$^\dagger$}
\footnotetext{Corresponding author. Email: svaikunt$@$uchicago.edu}
\endgroup


\section{Introduction}
Single cells in multicellular organisms perform complex tasks and remain functional in changing environments by adapting and maintaining cellular or tissue-level homeostasis. Examples of such phenomenology include the adaptation of force generation in airway smooth muscle cells in response to mechanical perturbation~\cite{gunst2003} and adaptation of cellular connectivity and cell division orientation in response to tension in developing tissues~\cite{nestor2019,guillot2013}.  
Such tasks demand sensing and processing of information about their environment and altering chemical or mechanical processes inside the cell to optimize cellular functions. 
Recent studies in single cells and tissues suggest that retention of memories of cellular morphologies and tuning of cellular features based on local cell environment may help their biological functionality ~\cite{arzash2025, tah2025,kalukula2024,fuhrmann2024}. This process is reminiscent of the concept of \textit{learning} wherein properties and behaviors are tuned such that multiple desired input output relations can be obtained. While mechanisms that enable learning are well studied in the neuronal context, it is not immediately clear if such mechanisms can be supported more broadly in cell mechanics.

In this work, we focus on a simplistic form of learning that aims to achieve a desired mechanical response. Our main results show how biomolecularly plausible force-sensitivities  - such as mechanosensitive proteins and force generator motors - allow systems to learn desired input-output force relations. Our work is built on the new paradigm of {\it physical learning}~\cite{stern2023rev,rocks2017, pashine2019, stern2021} and adapts recent work~\cite{falk2023} to show how learning may be possible with generic biomolecular ingredients present in a cell.

More specifically, we consider a simplified mechanical network caricature that could represent a coarse-grained version of the cellular cytoskeleton with essential regulatory actors - these mimic the various mechanosensitive proteins and motors that interact with the cytoskeleton -  and try to understand whether learning may emerge as a result of cell-autonomous mechanochemical interactions between these few key elements. Our main finding is that with particular kinds of mechnosensitivities, and rules for network remodelling due to stresses, our mechanical network caricature is able to learn (potentially multiple) input-output relations encoded in environmental signals. Building on and extending work in Ref.~\cite{falk2023}, we show that our cytoskeleton inspired mechanical networks can perform a version of \textit{contrastive learning}, a framework used routinely in machine learning. Importantly, we show how this learning persists even in cases with biophysical ingredients such as turnover of the constituent
elements of the network.

We also advance our analysis to explore if a cell-autonomous process can replace the external environment  and if a self-organized physical learning mechanism may emerge. Finally, to elucidate the role of physical learning in cellular functionalities, we explore how a version of our learning mechanisms may be vital for achieving the homeostasis of the mechanical state of the cell, e.g., homeostasis of tension or strain in the cytoskeletal structure.

\begin{figure}[t]
\begin{center}
    \includegraphics[width=0.95\columnwidth]{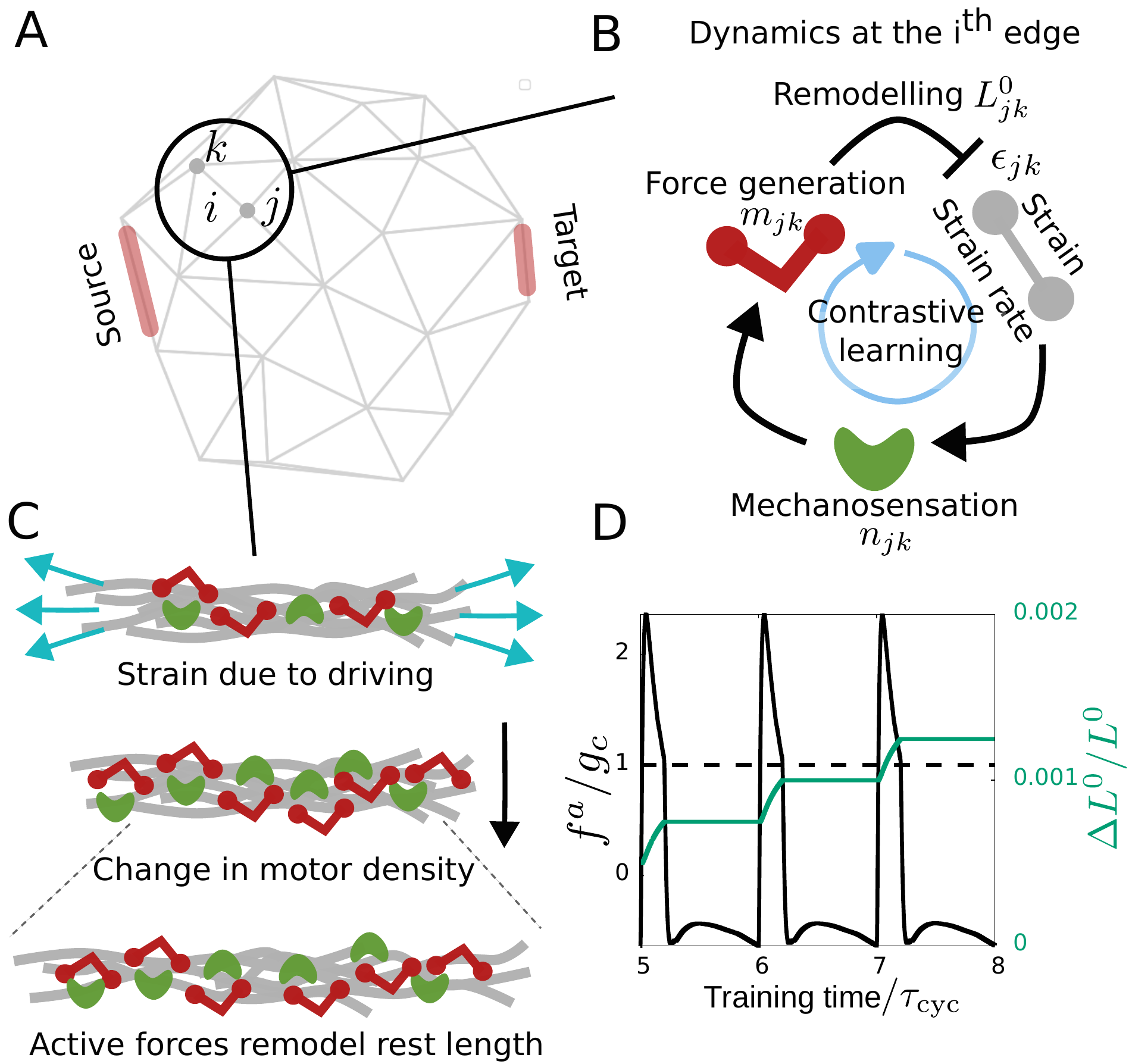}
    \caption{Physical learning via structural remodeling in a model inspired by the cytoskeleton:  (A) Our model is a disordered spring network of nodes connected by edges. The source and target edges are indicated in color. The learning mechanism involves supervised driving at the target edge. (B) Schematic representation of the molecular mechanism of learning showing the dynamic coupling between the network elasticity and agents mimicking molecular motors and mechanosensitive proteins that can bind/unbind from each edge. (C) Schematic showing how driving at the target edge creates local strain and enables an update of the learning degree of freedom (rest length) at any arbitrary edge of the network,  via mechanosensation and active force generation. (D) Learning dynamics in one edge of the network. The change in the learning degree of freedom (rest length) at the edge in response to active forces generated by motor dynamics. The dashed line indicates where the active force reaches the threshold active force value $f^a=g_c$ above which rest length changes according to the learning rule. }
    \label{fig:schematic}
\end{center}
\end{figure}



\section{Model}
 
We consider a disordered network inspired by the actomyosin cytoskeleton~\cite{linsmeier2016,broedersz2011,broedersz2014} (Fig.~\ref{fig:schematic}A). The edges of this disordered network can be conceived as cross-linked bundles of actin filaments that undergo structural remodeling based on local quantities such as active stress~\cite{hayakawa2001dynamic, noll2017, ruckerl2017adaptive, cavanaugh2020, staddon2019}. We shall consider the edges to be Hookean springs (i.e., linear elasticity) given by stiffness ($k$) and rest length ($L^0$). In addition, we consider the coupled dynamics of two molecular components: one that mimics an active stress-producing motor and one that mimics a mechanosensitive protein, both of which can bind/unbind from the edges. Although there are many types of molecular motors such as myosin, kinesin, etc., and different mechanosensitive proteins such as the LIM-domain proteins like zyxin, and paxillin, for simplicity here we consider two generic molecular components to capture the functional role of mechanosensitive LIM-domain proteins~\cite{anderson2021} and molecular motors. We describe the state of this cytoskeletal network by the position of the nodes $\{\mathbf{r}\}$, the number density of the molecular motor ($m$) and mechanosensitive protein ($n$) on the edges (Fig.~\ref{fig:schematic}B).

The length of the edge between the nodes $j$ and $k$ is given by $L_{jk}=|\boldsymbol{r}_j-\boldsymbol{r}_k|$ and determined by the dynamics of these two node positions. The node position changes according to overdamped dynamics and the equation for $j^{th}$ node is given by
\begin{equation}
    \gamma \dot{\boldsymbol{r}}_j = \sum_{k}^{nn} -k_{jk}(|\boldsymbol{r}_{j} - \boldsymbol{r}_{k}| - L^0_{jk}) \hat{\boldsymbol{r}}_{jk} + f^a_{jk} \hat{\boldsymbol{r}}_{jk} \, ,
 \label{eq:node}   
\end{equation}
where the sum is over the neighbors of node $j$ set by the network connectivity and the unit vector $\hat{\boldsymbol{r}}_{jk}$ denotes the direction from the $j^{th}$ node to $k^{th}$ node. $\gamma$ is the friction coefficient, $k_{jk}$ and $L^0_{jk}$ are the stiffness and rest length of the edge between the nodes $j$ and $k$. The active force in the edge due to motor contractility is given by $f_{jk}^a=\xi m_{jk}$ where $m_{jk}$ is the number density of the bound motors and $\xi$ is a positive constant known as the {\it contractility} or activity parameter. 
The motor binding-unbinding kinetics is known to be associated with mechanosensitive proteins~\cite {schiller2011}. The motor dynamics in the edge depends on the number density of mechanosensitive proteins, $n_{jk}$, and is given by 
\begin{equation}
    \dot{m}_{jk} = k^0_b + k^1_b n_{jk} - k_u m_{jk}, 
\label{eq:myosin}    
\end{equation}
where $k^0_b$ and $k_u$ are the bare binding and unbinding rates 
and $k^1_b$ is a constant factor that determines how these proteins promote motor binding.
Motivated by the tension-dependent recruitment of mechanosensitive proteins~\cite{lele2006,smith2014lim, schiller2011}, we consider force-dependent catch-bond-like dynamics for the number density of bound proteins where a higher strain rate leads to a lower unbinding rate. The number density of bound proteins in the edge between nodes $j$ and $k$ given by
\begin{equation}
  \dot{n}_{jk} = k_{bn} - k^0_{un} e^{-\beta \dot{\epsilon}_{jk}} n_{jk}\, ,
\label{eq:lim}
\end{equation}
where $k_{bn}$ and $k^0_{un}$ are the bare binding and unbinding rate. Here $\dot{\epsilon}_{jk}$ is the strain rate and the strain ($\epsilon_{jk}$) in the edge is defined as $\epsilon_{jk}=(L_{jk}-L_{jk}^0)$ and $\beta$ is the coefficient for strain-rate dependent unbinding. We linearize and rescale the above equations to derive a set of dynamical equations with unitless parameters as detailed in the Supplementary Section~\ref{appendix:model}.

We focus on a simple form of learning, namely, learning a desired strain response in mechanical networks~\cite{rocks2017, stern2021}. Learning, in the current context, occurs when a system changes its internal interactions between its components based on the system's response (output) to an external stimulus (input), such that its response to subsequent stimuli {\it improves} its ability to reach a desired state~\cite{stern2023rev}. We want the strain of a given magnitude in one part of the network (i.e. input at source) to create a desired amount of strain in a faraway part of the network (i.e. output at target). To build such an input-output relationship, the learning dynamics or training changes the interactions, that is, the rest length ($L^0$) and stiffness ($k$) of the edges based on the local active force ($f^a$). Thus, the rest length and stiffness of the edges become the learning degree of freedom (LDOF) governed by the learning dynamics or the {\it learning rule}.
In the following, we show how this setup naturally allows our system to learn and remodel in response to environmental stimuli. In particular, adapting Ref~\cite{falk2023}, we argue that the mechanosensitive proteins and remodeling by active forces allow our network to implement a contrastive learning protocol - a common learning framework in the machine learning context.

\section{Feedback between mechanosensitive proteins and molecular motors can provide a pathway for learning}

To see how this setup can enable the learning and encoding of mechanical response via external signals, we begin by considering the coupling of the mechanosensitive protein ($n$) and motor ($m$) dynamics. The strain rate dependence of the mechanosensitive protein kinetics enables the local motor dynamics to have a history of local strain, providing a way for biologically plausible contrastive learning~\cite{falk2023}. 
Using the linearized form of the above equations Eq.~\ref{eq:myosin}-\ref{eq:lim}, we can show that the motor dynamics possesses the memory of the local strain in the form of,
\begin{eqnarray}
    \delta m &=&\int^t_{-\infty} \mathcal{K}(t-t') \delta\epsilon(t')  dt',
\label{eq:myo_memory}    
\end{eqnarray}
where the memory kernel $\mathcal{K}$ is given by 
\begin{equation}
    \mathcal{K}(t-t')=\beta_1 \left( \delta(t-t') - \frac{1}{\tau_k} e^{-\frac{(t-t')}{\tau_k} } \right) \, ,
    \label{eq:kernel}
\end{equation}
where $\delta m$ and $\delta\epsilon$ are the variation of motor density and strain around a steady state given by $m^0$ and $\epsilon^0$  (see Appendix~\ref{appendix:model} for details).
Using the consistency between the integral form (Eq.~\ref{eq:myo_memory}) and the motor dynamics described in Eq.~\ref{eq:myosin} we can identify the kernel timescale $\tau_k$ and the coefficient $\beta_1$ to be $k_u^{-1}$ and $k^1_b \beta n_0$ respectively (see Appendix~\ref{appendix:memory} for details). It is evident from the memory kernel (Eq.~\ref{eq:kernel}) and the form of the constant $\beta_1=k^1_b n_0 \beta$ that the strain-rate-dependent mechanosensitive protein dynamics ($\beta \ne 0$) and its coupling with the motor dynamics ($k^1_b \ne 0$) are crucial for the local motor dynamics to have the memory of the local strain. We shall use the value of $\beta_1$ as a measure of mechanosensitivity later in this study.

 Recent work by Falk {\it et al}~\cite{falk2023}
shows that implicit memory of the quantity of interest (e.g., memory of local strain in the current case as given by Eq.~\ref{eq:myo_memory}, \ref{eq:kernel}) can enable learning through a mechanism commonly referred to as contrastive learning. 
In the most generic implementation of this paradigm, the LDOF of the system are updated by comparing the quantity of interest in the {\it free} and the {\it clamped} state which correspond to the native and desired state of the system after applying the input. According to Ref~\cite{falk2023}, if the system is exposed to a signal that periodically oscillates between imposing a constraint on the output nodes that nudges it towards a desired value ---
while the input nodes are held fixed at a specific value (i.e., \textit{clamped}) --- and being \textit{free}, the system can remodel through a contrastive learning mechanism so that it learns the desired responses. These arguments, specialized for now to the linear limit, show how force-sensitive kinetics can enable a learning mechanism. Importantly, in the following, we show that this learning mechanism works broadly, even with non-linear effects and biophysical effects such as turnover, and not just in the linear limit. 

\section{Dynamics of training and learning}

We apply the dynamical procedure described in the previous section to our system, driving it periodically to shift between the free and the clamped states by applying an external force at the target.
The system is taken from the free state to the clamped state very fast over a timescale $\tau_f$ and slowly brought back over a longer timescale $\tau_s$ (Supplementary Fig.~\ref{fig:model_SI}). Training involves driving the system through many such cycles, that is, the time duration of one cycle is $\tau_\text{cyc}=\tau_f + \tau_s$. We shall discuss the training process in more detail in the following sections.

The external drive and implicit memory combine to train the system by changing the learning degrees of freedom via a learning rule. Based on the structural remodeling dependent on the active force in the network edges, we shall consider the dynamics of the LDOF, i.e. the rest length ($L^0_{jk}$) and the stiffness ($k_{jk}$) of the edges depending on the local active force $f^a_{jk}$ in the edge. First, we shall consider unalterable edge stiffness and only the rest length to be the LDOF with the dynamics given by
\begin{equation}
    \dot{L}^0_{jk} = \alpha g(f^a_{jk}) \, ,
\end{equation}
where $\alpha$ is the learning rate parameter and $g(x)=x$ for all $|x|\ge g_c$ and $0$ otherwise. The parameter $g_c$ represents a threshold active force above which the edges remodel their rest length. The active force being proportional to the motor density which possesses the implicit memory of strain, the above learning rule results in contrastive update of the LDOF (rest length). 
While we introduce the learning mechanism considering linear elasticity of the network without turnover of edges and linear approximation in the mechanosensitive protein dynamics (Eq.~\ref{eq:lim}), we shall discuss how the learning mechanism is affected by nonlinearities and network turnover later in this work.

\section{Learning strain response in a model cytoskeletal network}\label{sec:allosteric_lrn}

Using the dynamics described above, we explore whether the network can learn a specific desired mechanical response. In particular, we consider two edges ---one source edge and one target edge (Fig.~\ref{fig:schematic}A) --- and train the network to learn to contract to a specific strain value $\epsilon^*_T$ at the target edge when we apply an extension of $\epsilon^*_S$ at the source edge. To train the network to produce the desired strain $\epsilon^*_T$ at the target edge, we first apply the strain $\epsilon^*_S$ at the source edge which takes the system into the {\it free state}. Then, on the target edge, we apply an external supervised force $\boldsymbol{f}^e$ which is proportional to the difference between the strain value at the target edge in the free state $\epsilon_T$ and the desired strain value at the target edge $\epsilon^*_T$, and is given by
\begin{equation}
  \boldsymbol{f}^e = -\lambda(t)\boldsymbol{\nabla} (\frac{1}{2}|\epsilon_T - \epsilon_T^*|^2 )\, ,
\label{eq:driving}
\end{equation}
where $\lambda(t)$ controls the temporal dynamics of forcing. It goes from $\lambda=0$ where the system is in the free state to $\lambda=\lambda_\text{max}$ where the system is in the clamped state over the fast timescale $\tau_f$ and returns to $\lambda=0$ over the slow timescale $\tau_s$ (Supplementary Fig.~\ref{fig:model_SI}). This asymmetry in the timescale of driving (i.e., $\tau_f<\tau_s$) is a hyperparameter which in previous work was shown to be necessary for successful learning~\cite{falk2023}. Due to this driving, the force balance equation on the target nodes (nodes connected by the target edge) becomes
\begin{equation}
    \gamma \dot{\boldsymbol{r}}_j = \sum_{k}^{nn} -k_{jk}(|\boldsymbol{r}_{j} - \boldsymbol{r}_{k}| - L^0_{jk}) \hat{\boldsymbol{r}}_{jk} + 
    \xi m_{jk}
    \hat{\boldsymbol{r}}_{jk} + \boldsymbol{f}^e \,.
 \label{eq:target_node}   
\end{equation}


\begin{figure}[ht!]
\begin{center}
    \includegraphics[width=0.99\columnwidth]{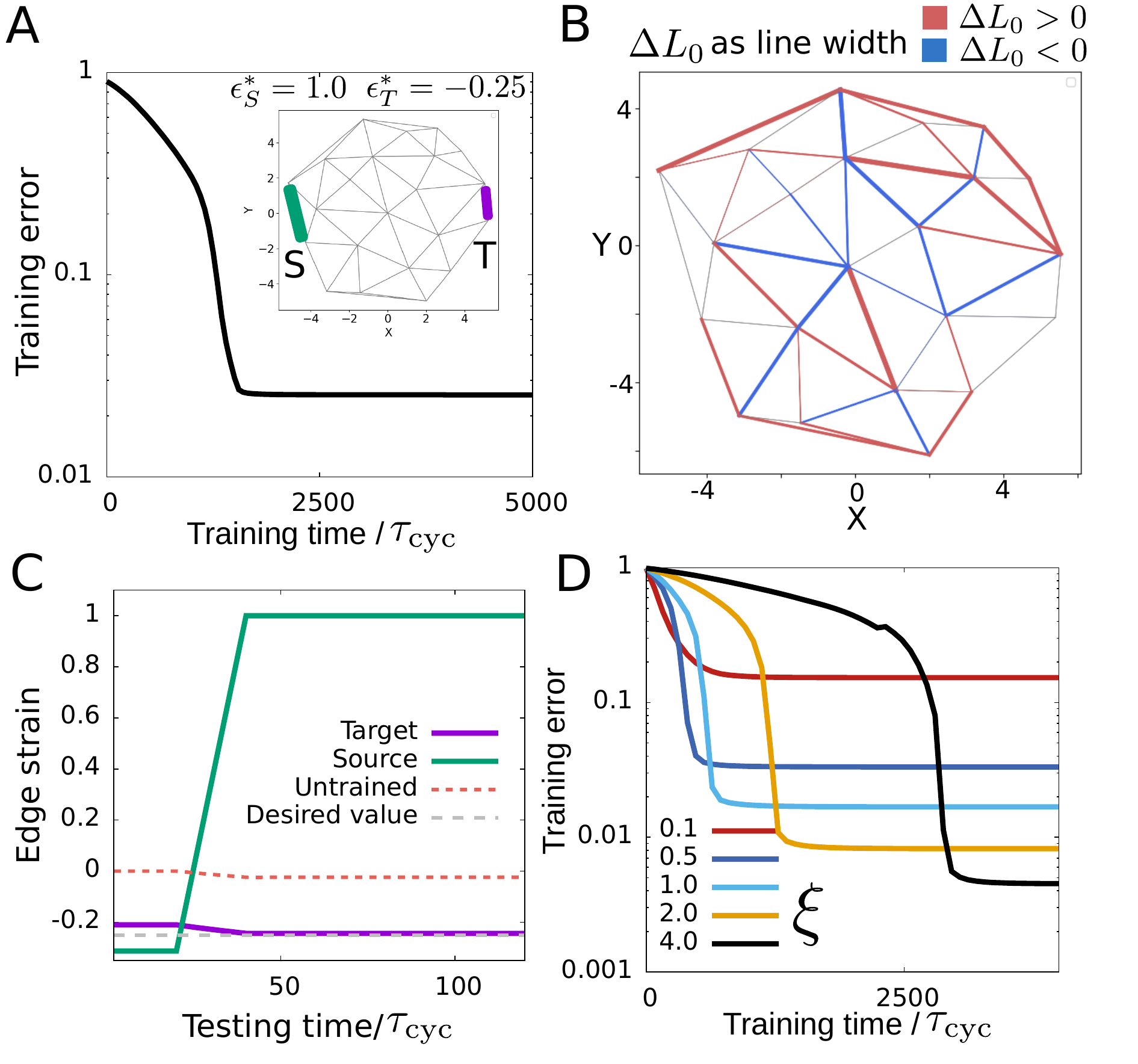}
    \caption{Learning strain response in cytoskeletal networks by changing rest lengths. (A) Time evolution of training error over many cycles of exposure to desired behavior. (inset) The network at the initial unstrained state. The source and target edges are denoted by ``S" and ``T" respectively. (B) Trained network reflecting the magnitude of changes in the rest length ($|\Delta L_0|$) of the edges given by the edge thickness. The colors indicate if the rest length increased (red) or decreased (blue) with respect to the original rest length values after training.  (C) The strain at the target edge on the application of the source strain on the trained network. The untrained response at the target edge is shown in a red dashed line which is far from the desired target strain (gray dashed line) (D) Training error vs training time at different values of activity given by the contractility parameter $\xi$. Here the rescaled parameter values are $k=1$, $\xi=0.5$, $\beta=0.1$, $k_u=0.5$, $\lambda_{\rm max}=0.5$ and $\alpha=1$.}
    \label{fig:strain_response_l0}
\end{center}
\end{figure}


Driving at the target edge changes the density of the mechanosensitive proteins and motors and subsequently the strain at other edges in the network (Eq.~\ref{eq:myosin},~\ref{eq:lim} \& Fig.~\ref{fig:schematic}C). As indicated previously, we expect that due to the implicit memory (Eq.~\ref{eq:myo_memory}), the motor dynamics at any edge ($i$) estimates the derivative of the local strain, which in turn has the information of the difference in local strain in the clamped and free states $\sim \epsilon^\text{clamped}_i-\epsilon^\text{free}_i$  because of the temporal driving at the target edge which periodically takes the system from the free to clamped state and back.
Thus, this information gets fed back in the local active force which is proportional to the bound motor density and the active force-dependent remodeling of rest length facilitates contrastive learning (Fig.~\ref{fig:schematic}D). The training error, defined as $|(\epsilon_T - \epsilon^*_T)|/\epsilon^*_T$ shows the effectiveness of the contrastive learning mechanism as it reaches close to zero as the training progresses (Fig.~\ref{fig:strain_response_l0}A). The spatial distribution of increase and decrease in the rest length values ($\Delta L_0$) over the network shows the trained network with altered interaction that led to this learned mechanical response (Fig.~\ref{fig:strain_response_l0}B). The trained network shows a distinctly different strain value which is close to the desired value $\epsilon^*_T$ at the target edge compared to the un-trained network (Fig.~\ref{fig:strain_response_l0}C). The learning dynamics remodels the rest lengths of various edges to achieve the desired strain at the target edge. This leads to a trained network that is appropriately pre-strained (i.e., with non-zero edge strain, $\epsilon_{jk}= L_{jk}-L_{jk}^0$, in equilibrium without any source strain) to achieve the desired mechanical response (Supplementary Fig.~\ref{fig:SI_trained_net}).
We repeat the learning problem in randomly chosen target and source edges on the periphery of the network to find the average training error to reduce over time indicating the learning capability to be independent of the particular choice of source and target edges (Supplementary Fig.~\ref{fig:SI_system_size}). We do not find any significant effect of the network size on learning capability (Supplementary Fig.~\ref{fig:SI_system_size}). The contractility parameter, $\xi$ ($f^a_{jk}=\xi m_{jk}$) gives the extent of activity in the network. We find the learning to become better with increasing activity ($\xi$) reaching lower training error values as the activity increases (Fig.~\ref{fig:strain_response_l0}D). Depending on various hyper-parameter values such as threshold active force ($g_c$) and timescales of driving ($\tau_f, \, \tau_s$) the learning may slowdown with increasing activity (Fig.~\ref{fig:strain_response_l0}D) which stems from increased unlearning (reversal of the change in rest length) in each cycle during training (Supplementary Fig.~\ref{fig:SI_unlearning}).

\begin{figure}[ht!]
\begin{center}
    \includegraphics[width=0.99\columnwidth]{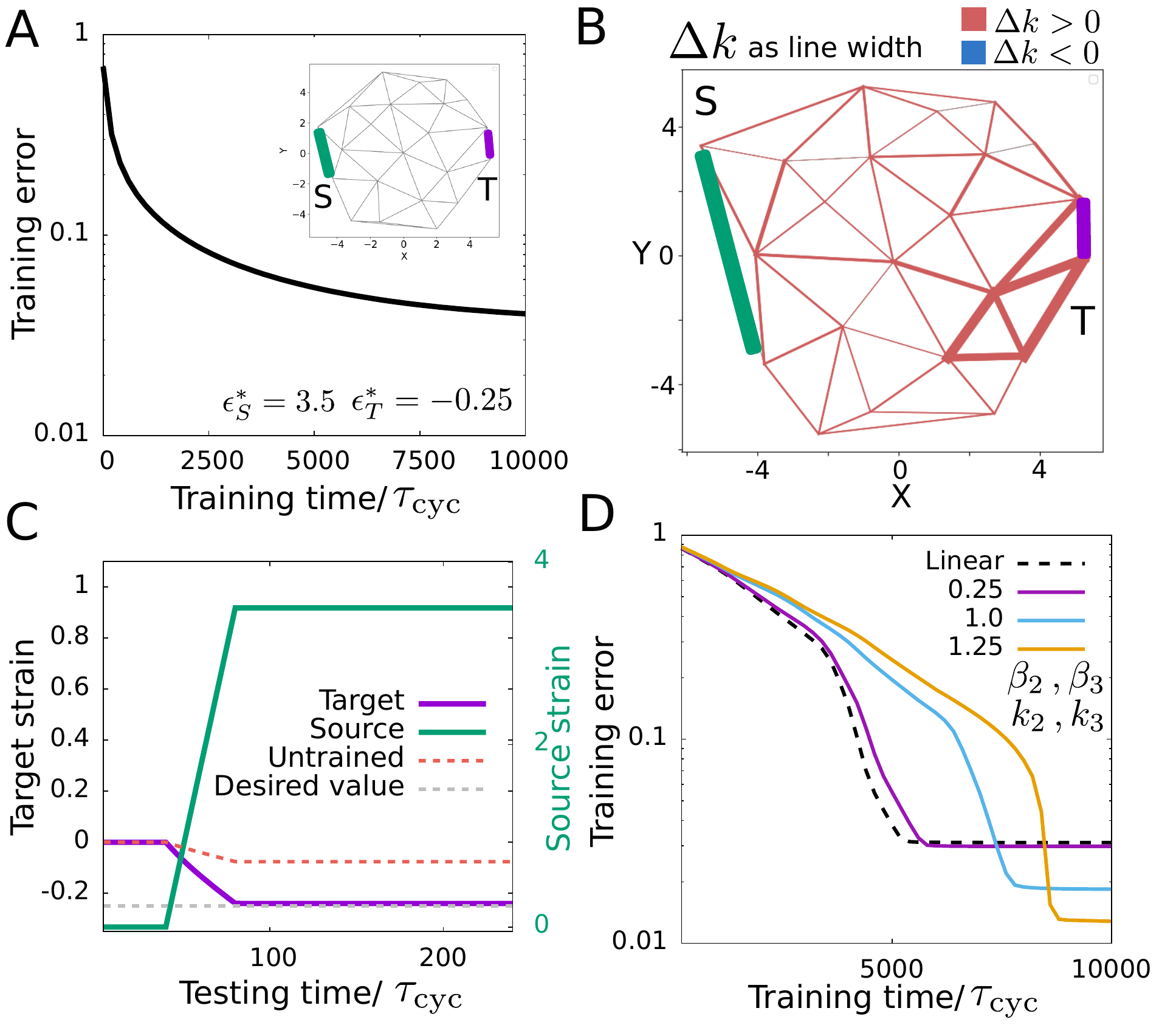}
    \caption{Learning strain response in cytoskeletal networks by changing stiffness. (A) Time evolution of training error over many cycles of exposure to desired behavior. (B) The trained network topology and edge stiffness change $|\Delta k|$ which is proportional to line width. The source and target edges are denoted by ``S" and ``T" respectively. (C) The strain at the target edge on the application of the source strain in the trained network. The untrained response at the target edge is shown in a red dashed line, which is far from the desired target strain (gray dashed line). (D) Learning with non-linear protein dynamics and non-linear elasticity. Training error vs training time shows that increasing the strength of nonlinearity in both the protein dynamics ($\beta_2,\beta_3\ne 0$) and network elasticity ($k_2,k_3\ne 0$) leads to lower training errors. The numbers in the legend indicate these parameter values. We consider a simple case of the same values for the unitless nonlinear coefficients $\beta_{2}=\beta_{3}=0.1, \, k_2=-0.1\,\& \,k_{3}=0.1$ with other parameters same as Fig.~\ref{fig:strain_response_l0} except $\lambda_{max}=0.2$ and $\xi=0.2$. For panels (A-C) the rescaled parameter values are $\xi=5$, $\beta=0.1$, $k_u=0.5$, $\lambda_{\rm max}=10$  and $\alpha=10^3$.}
    \label{fig:strain_response_k}
\end{center}
\end{figure}


We arrive at a different type of {\it solution} for the same learning problem (i.e., as described at the beginning of this section) if we consider active force-dependent remodeling of the edge stiffness values while the rest length values remain unaltered. In that case, the dynamics of edge stiffness depends on $\epsilon f^a$ and for $i^{th}$ edge, is given by
\begin{equation}
    \dot{k}_{jk}=\alpha g(\epsilon_{jk} f^a_{jk}) \, .
\end{equation}
Similar to the previous case, the driving at the target edge affects local motor density and strain values at each edge and that in turn changes the stiffness incrementally in every switching from free to clamped state. The training error reduces as the edges change their stiffness and the network learns the desired target strain value (Fig.~\ref{fig:strain_response_k}A) and the network develops a spatial pattern of heterogeneous but always positive stiffness value changes over the network, with the largest increase close to the target (Fig.~\ref{fig:strain_response_k}B). Here, the trained network achieves the desired mechanical response which is distinctly different from the untrained response (Fig.~\ref{fig:strain_response_k}C). In contrast to the previous case, here the trained network does not have any pre-strain (Supplementary Fig.~\ref{fig:SI_trained_net}). Rather, it reaches the correct solution by changing how the target edge contracts in response to the source edge extension (Fig.~\ref{fig:strain_response_k}C).


In the above results, we use a linear approximation in the dynamics of the mechanosensitive proteins (Eq.~\ref{eq:lim}) $e^{-\beta \dot{\epsilon}_{jk}}\simeq 1-\beta \dot{\epsilon}_{jk}$. We find no significant qualitative or quantitative changes in the learning dynamics (i.e., training error evolution over time) when we consider nonlinearities in the strain-rate dependent unbinding term in the protein dynamics while learning a desired mechanical response by changing rest length of the edges, same as described earlier in this section. We consider higher-order terms up to $3^{rd}$ order (i.e., 
$e^{-\beta \dot{\epsilon}_{jk}}\simeq 1-\beta \dot{\epsilon}_{jk}+\frac{\beta^2}{2} \dot{\epsilon}^2_{jk}-\frac{\beta^3}{6} \dot{\epsilon}^3_{jk}$) which changes the implicit memory the motor dynamics contains compared to the linear case. With the nonlinearity, the motor dynamics becomes dependent on a nonlinear function of strain rate $\delta \dot{m} \sim \beta_1 \delta \dot{\epsilon} - \beta_{2} \delta \dot{\epsilon}^2 + \beta_{3} \delta \dot{\epsilon}^3$ (see Appendix~\ref{appendix:nonlinear} for details). Despite this imperfection in implicit memory, which is not simply the derivative of local strain at each edge and does not estimate the difference in local strain between the free and clamped states, the learning capacity of the network remains largely unaffected for moderate strength of the nonlinear terms (Fig.~\ref{fig:nonlinear_SI}A). This indicates that the linear approximation is adequate for describing the learning mechanism.

Apart from the nonlinearity in the mechanosensitive protein dynamics, one could also consider the nonlinear elastic response of the network, e.g., strain stiffening. Cytoskeletal networks are known to have nonlinear elastic properties~\cite{xu2000, janmey1994}. We consider a network with nonlinear elasticity where the energy as a function of edge strain($\epsilon$) is given by: $E_{el}=\frac{1}{2} k\epsilon^2 + \frac{1}{3} k_{2} \epsilon^3 + \frac{1}{4} k_{3} \epsilon^4$ to see the effect of nonlinear elasticity on learning. Here, $k_{2}$ and $k_{3}$ are the elastic constants for higher-order terms in the elastic energy. For a network with nonlinear elasticity, the nonlinearities in the mechanosensitive protein dynamics (up to 3rd order) lead to the correct implicit memory required for the temporal contrastive learning (see Appendix.~\ref{appendix:nonlinear} for details). Thus, nonlinearity in the protein dynamics enables contrastive learning in a network with nonlinear elasticity. Interestingly, we find lower training error with increasing strength of the nonlinearity indicating nonlinear elasticity can aid learning (Fig.~\ref{fig:strain_response_k}D).

\section{Classification of environmental signals}

\begin{figure}[t]
\begin{center}
    \includegraphics[width=1.02\columnwidth]{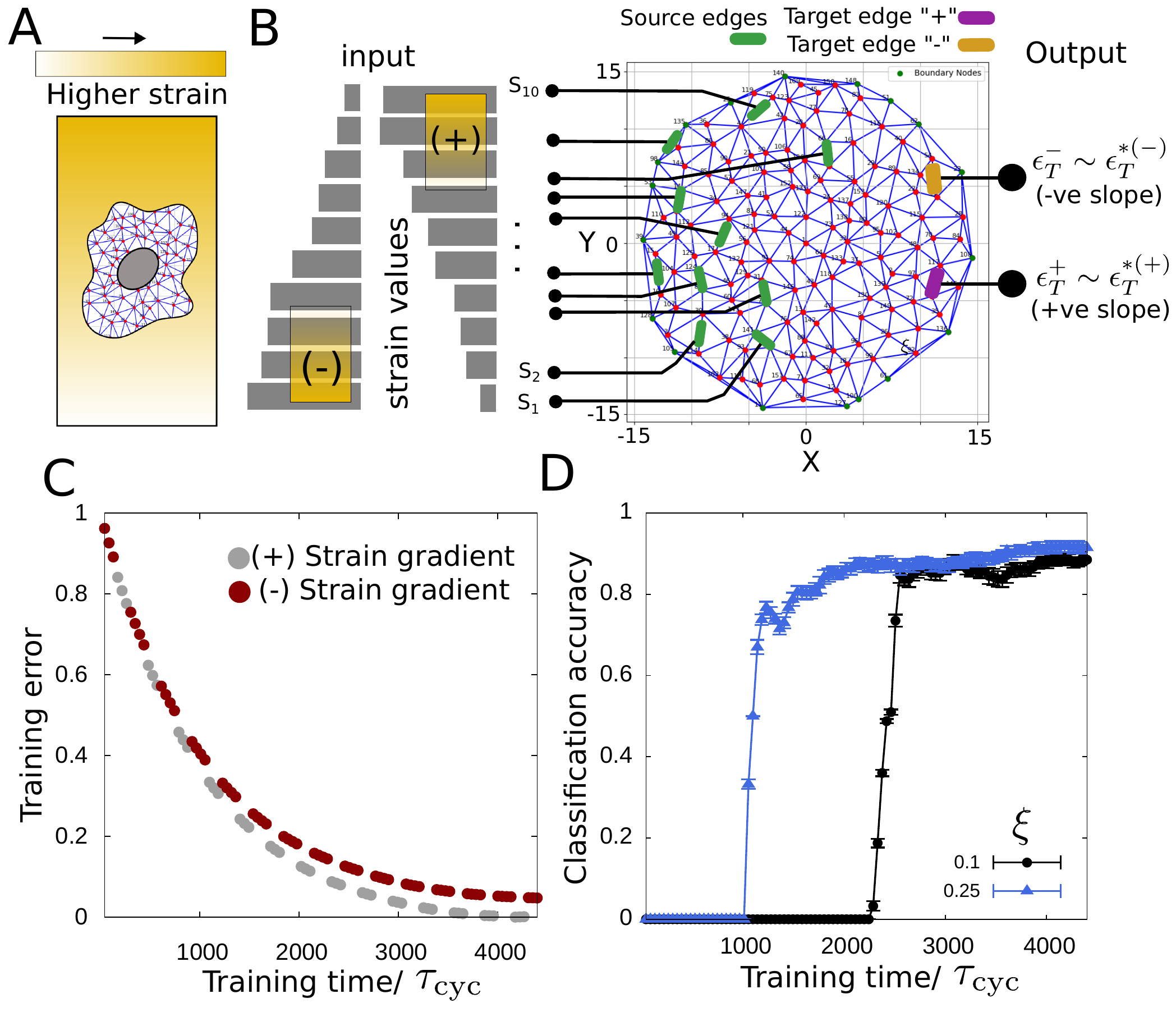}
    \caption{Classification of environmental signals in cytoskeletal networks. (A) Schematic showing a cell in an environment with a strain gradient. (B) Environmental mechanical signal as the positive and negative gradient of strain. The strain pattern input is imposed on source edges (green) and two target edges are chosen to encode the two classes $``+"$ (purple) and $``-"$ (orange). (C) Training error vs training time for both positive and negative strain gradients as inputs. (D) Classification accuracy increases as the training progresses. The network learns to classify faster with higher activity (given by contractility parameter $\xi$). Here $\tau_f/\tau_s=1/10$, $\epsilon^{*(+)}_T=-\epsilon^{*(-)}_T=0.25$, $\mathcal{E}_{tol}=0.025$ and the rescaled parameter values are same as used in Fig.~\ref{fig:strain_response_l0} except $\alpha=10$.}
    \label{fig:contrast_class}
\end{center}
\end{figure}

Another classic learning task is solving a classification problem via learning to distinguish between input signals of different types (classes) and produce different responses as output by creating a map between these sets of input and output signals. 
Living cells continuously interpret noisy environmental signals to decide the future course of their action~\cite{perkins2009}. For example, both chemical and mechanical environmental signals are known to affect stem cell differentiation decisions~\cite{discher2009}. Synthetic chemical reaction networks that can classify chemical signals within the cell have recently been reported~\cite{chen2024}.
Here we consider a task where the cytoskeletal network is trained to differentiate between different mechanical signals, in particular, different strain gradients in the environment (Fig.~\ref{fig:contrast_class}A). Deciphering such environmental mechanical signals may be important for cellular functions such as cell motility~\cite{yang2024} and tissue remodeling during development~\cite{reig2017}.


We consider increasing and decreasing strain patterns (i.e., positive and negative strain gradients) as two distinct types of environmental mechanical signals. This strain gradient is relative to the source edge indices chosen according to their spatial position along the y-axis (Fig.~\ref{fig:contrast_class}B). The slopes of the gradients are drawn from two uniform distributions $\mathcal{U}_1\in (0.01,0.1)$ and $\mathcal{U}_2\in (-0.1,-0.01)$. We consider $10$ source edges ($S_1, S_2, \ldots, S_{10}$) in the network and two target edges, target edge ``$+$" and target edge ``$-$" to encode the information of the positive and negative gradients as the desired target strain values $\epsilon^{*(+)}_T$ at the ``$+$" and $\epsilon^{*(-)}_T$ at the ``$-$" target edge correspondingly via training. 
During training, one slope value ($s$) was drawn randomly from either $\mathcal{U}_1$ or $\mathcal{U}_2$ in each switching cycle (i.e., a time duration of $\tau_s + \tau_f$) and depending on the sign of the slope, either the ``$+$" or the ``$-$" target edge was trained to reach the corresponding target strain values. 
The source strain of the $n^{th}$ source edge is given by $\epsilon^*_{S(n)}=n s r$ for the positive slope of strain and  $\epsilon^*_{S(n)}=1-(n-1)s r$ for the negative slope of strain with $r$ being a uniform random number used to introduce small noise (up to $20\%$, i.e., $r\in [0,0.2]$) in the source strain values. 
The training error for both the positive ($|\frac{\epsilon_T^{(+)} - \epsilon^{*(+)}_T}{\epsilon^{*(+)}_T}|$) and negative ($|\frac{\epsilon_T^{(-)} - \epsilon^{*(-)}_T}{\epsilon^{*(-)}_T}|$) input strain-gradients reduces as the training progresses (Fig.~\ref{fig:contrast_class}C).
In the trained network, the target strain values obtained in response to the imposed source strain pattern will be used to evaluate the performance of the classification task. For example, if the target edge ``$+$" reaches the closest to the desired strain value $\epsilon_T^{*(+)}$ and the difference is smaller than a tolerance value ($\mathcal{E}_{tol}$), i.e., $|\epsilon_T^{(+)}-\epsilon_T^{*(+)}|<\mathcal{E}_{tol}$ and $|\epsilon_T^{(+)}-\epsilon_T^{*(+)}|<|\epsilon_T^{(-)} - \epsilon_T^{*(-)}|$ when a positive strain gradient is presented in the sources, the classification of that strain signal by the network is considered successful. We define classification accuracy as the ratio of the number of successful classifications to the number of imposed source strain gradients, i.e., the number of tests. 

A set of strain gradients was drawn from $\mathcal{U}_1$ and $\mathcal{U}_2$ and the classification output was checked as we trained the network.
The behavior of the trained network indicates the successful classification of the two classes of environmental signals as the classification accuracy increases with the progression of training (Fig.~\ref{fig:contrast_class}D). Since increasing the network activity, as measured by the connectivity parameter $\xi$, leads to lower training error, we find increasing activity to result in faster classification in the network (Fig.~\ref{fig:contrast_class}D).

\section{Learning in the presence of turnover of network components}

\begin{figure}[t]
\begin{center}
    \includegraphics[width=1.0\columnwidth]{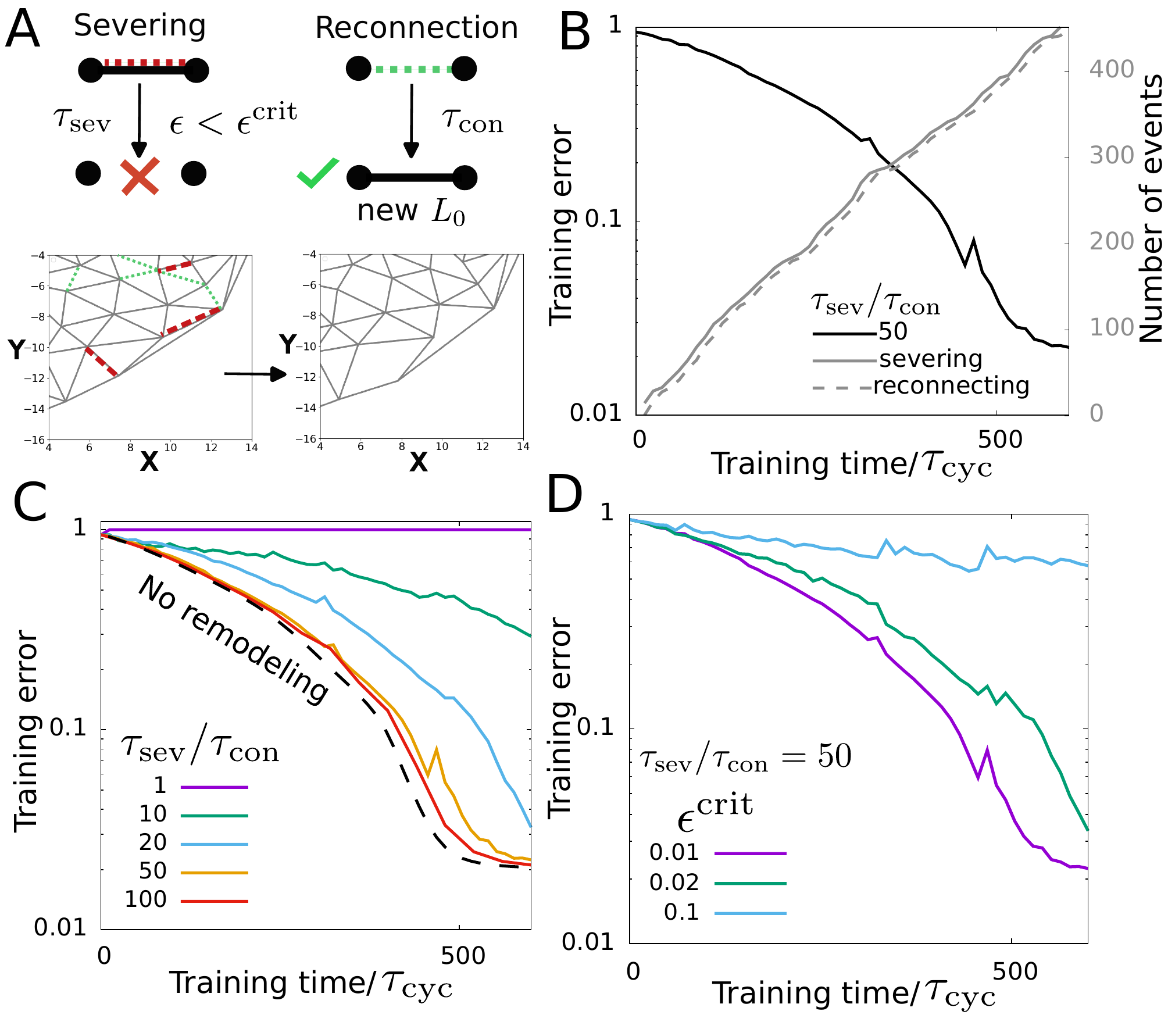}
    \caption{Learning in the presence of network turnover. (A) The schematic on the top shows edge severing and reconnection processes. The bottom part shows an example of network remodeling. Green and red dashed lines mark the bonds to get created and severed respectively. (B) Training in the presence of network remodeling. Training error decreases over time. The solid and the dashed gray lines show the number of severing and reconnection events as the training progresses. (C)  Training error as a function of the training time at various severing timescales ($\tau_\text{sev}$). The dashed line shows training without remodeling. (D) Training error with training time at different critical strain values $\epsilon^\text{crit}$. $\tau_\text{sev}/\tau_\text{con}=50$. The parameter values used here are the same as Fig.~\ref{fig:strain_response_l0} except $\epsilon^{\rm crit}=0.01$, $\tau_{\rm con}=10\tau_{\rm cyc}$.}
    \label{fig:remodeling}
\end{center}
\end{figure}

Cellular cytoskeletal networks undergo continuous turnover of their components. To understand how learning dynamics is affected by the turnover of network components, we introduce dynamic turnover of the network edges during training. 
The edge turnover dynamics is implemented by severing of edges with a timescale $\tau_\text{sev}$ and reconnecting the nodes of the edge severed with a timescale $\tau_\text{con}$. During reconnecting, the instantaneous distance between the nodes is set as the new rest length of the reconnected edge. Additionally, edges with strain higher than a critical value $\epsilon > \epsilon^\text{crit}$ do not sever (Fig.~\ref{fig:remodeling}A). This strain dependence of the severing captures catch-bond-like dynamics seen in the cytoskeletal networks, where parts of the network with higher tension are more stable~\cite{hayakawa2011, sato2005}. Thus, in each update during training (i.e., from $t\to t+ dt$), the edges with strain smaller than the critical strain ($\epsilon < \epsilon^\text{crit}$) sever with a probability given by $dt/\tau_\text{sev}$. Similarly to severing, reconnection of the severed edges occurs with a probability $dt/\tau_\text{con}$. 

We train the network to achieve a desired mechanical response using the rest length of the edges to be the LDOF, the same as described in Section.~\ref{sec:allosteric_lrn}. To avoid trivial errors, the source and target edges were excluded from the severing and reconnecting dynamics. We find the learning to be successful with a significant number of severing and reconnecting events at the edges (Fig.~\ref{fig:remodeling}B) which leads to a large number of edges being severed during training (Fig.~\ref{fig:SI_remodeling}). 
Learning becomes slower and the ability to learn decreases as the severing timescale becomes smaller and similar to the timescale of reconnecting (Fig.~\ref{fig:remodeling}C).

The learning altogether stops (i.e., target edge strain cannot reach the desired value) at small values of the ratio of the severing timescale and the reconnection timescale $\tau_\text{sev}/\tau_\text{con}\lesssim 10$. Note that we consider severing and reconnection timescales larger than the timescale of driving (i.e., $\tau_\text{sev},\,\tau_\text{con} > \tau_\text{cyc}$) and faster turnover dynamics would result in loss of learning even at high $\tau_\text{sev}/\tau_\text{con}$ values. 
The connectivity, defined as the average number of neighbors of a node, of the dynamic network decreases as the severing time scale decreases, that is, the severing rate increases, which may lead to the loss of learning ability seen at low severing timescales (Fig.~\ref{fig:SI_remodeling}). 
The choice of the critical strain value ($\epsilon^\text{crit}$) also significantly affects learning and leads to a slowdown of learning with increasing $\epsilon^\text{crit}$ and loss of learning ability for high $\epsilon^\text{crit}$ values (Fig.~\ref{fig:remodeling}D). This loss of learning may be the result of the loss of edges that propagate tension from the target edge to the network, getting remodeled.

\begin{figure}[t]
\begin{center}
    \includegraphics[width=1.02\columnwidth]{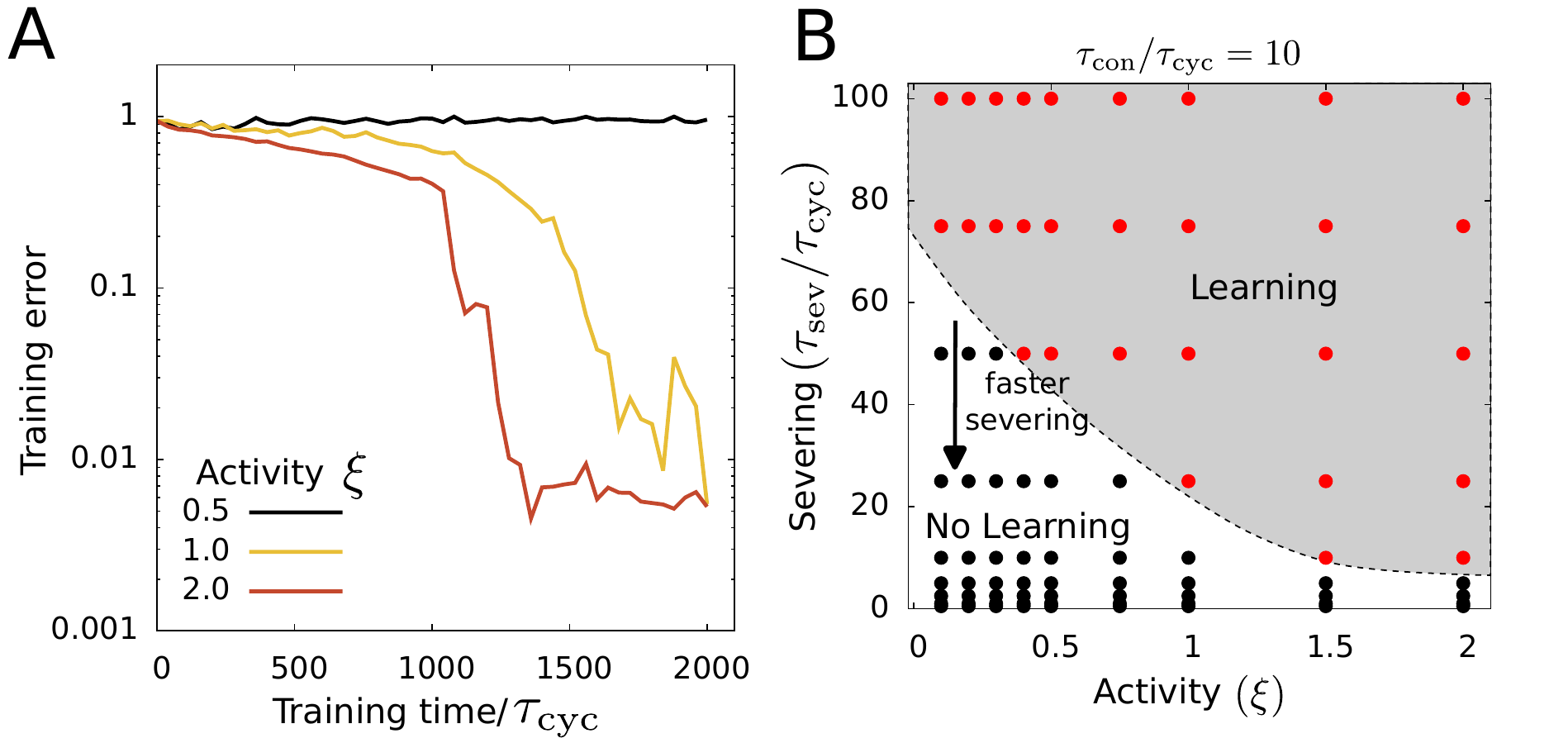}
    \caption{Activity dependent recovery from loss of learning. (A) Training error for different activity values (given by the contractility parameter $\xi$) for $\tau_\text{sev}/\tau_\text{con}=9$. (B) A phase diagram in activity and rescaled severing timescale. The arrow indicates the direction of the faster turnover of network edges. The parameter values used here are the same as Fig.~\ref{fig:strain_response_l0}. Additionally $\epsilon^{\rm crit}=0.01$, $\tau_{\rm con}=10\tau_{\rm cyc}$.}
    \label{fig:remodeling_phasedia}
\end{center}
\end{figure}

Interestingly, we find that the network can recover from the loss of learning with increased activity (Fig.~\ref{fig:remodeling_phasedia}A). At a fixed $\epsilon^\text{crit}$ value, increased activity reduces the probability of severing the edges by increasing the strain of the edges, which leads to the recovery of learning. The ability of the network to learn in the presence of edge turnover depends on the balance between the severing dynamics and the activity of the network. The network learns better at higher activity values (large $\xi$) and slower severing rates (larger $\tau_\text{sev}$) while low activity and fast turnover lead to loss of learning in the network (Fig.~\ref{fig:remodeling_phasedia}B).  

\section{Learning via self-organized pulsation}
\begin{figure}[t]
\begin{center}
    \includegraphics[width=0.95\columnwidth]{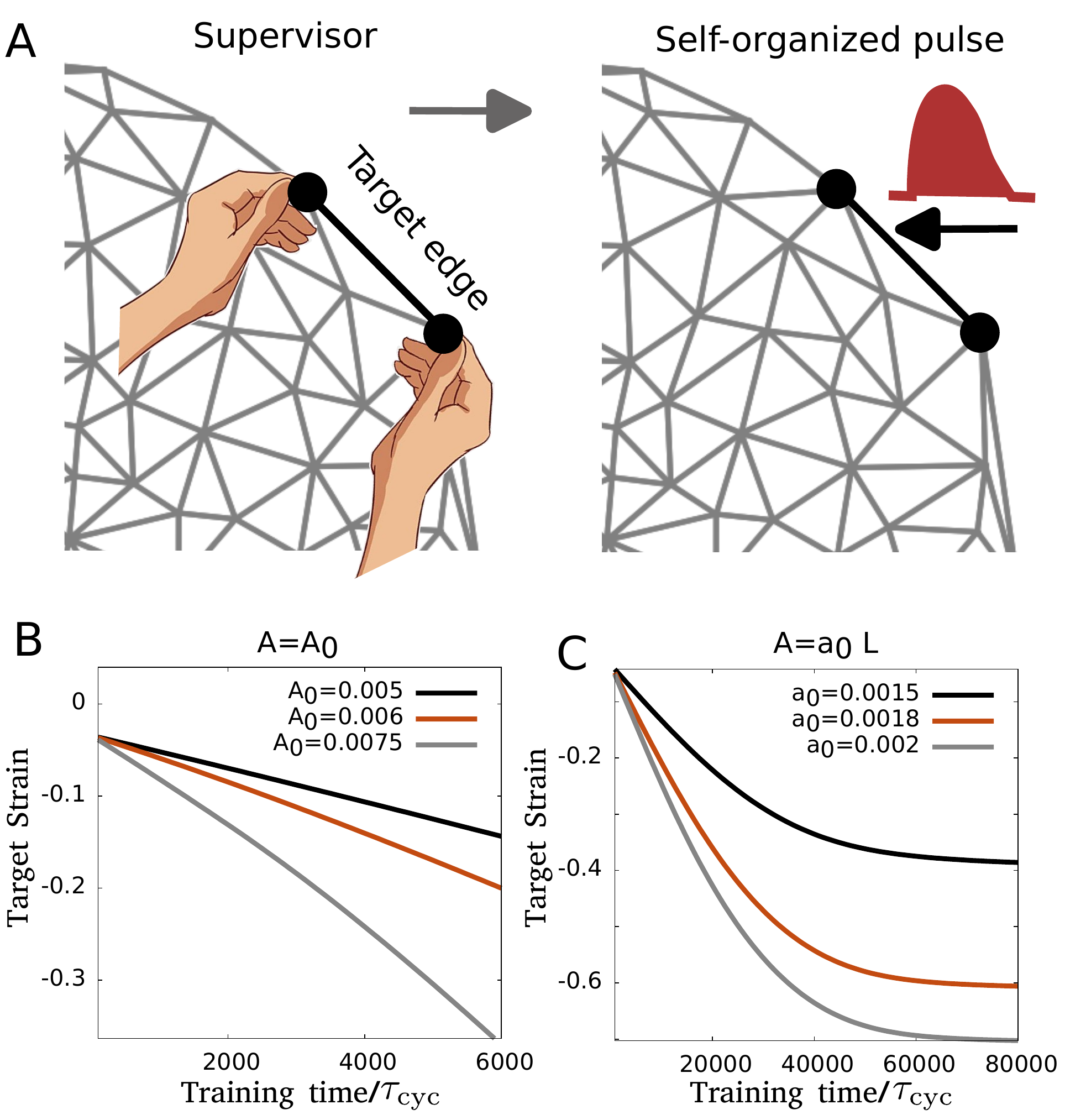}
    \caption{Self-organized learning via actomyosin pulsation. (A) Schematic showing self-organized pulse driving the target edge instead of the supervisor.
    (B) Target strain with training time for driving by pulses of constant amplitude $A=A_0$. (C) Target strain with training time for driving by pulses with target edge length ($L$) dependent feedback $A=a_0 L$. Here $\tau_f/\tau_s=1/4$ and the rescaled parameter values are $\beta=0.1$, $k_u=0.4$, $\xi=1$ and $\alpha=1$. }
    \label{fig:pulse_lrn}
\end{center}
\end{figure}

The learning mechanism that we have discussed so far requires an external drive at the target edge. This driving requires knowledge of the desired target strain $\epsilon^*_T$ (Eq.~\ref{eq:driving}) that determines the clamped state. In all of the above cases the training was performed with a fictional ``supervisor" who knows the value of $\epsilon^*_T$ playing the crucial role of driving the system from the free to clamped state (Fig.~\ref{fig:pulse_lrn}A), thus making the above learning mechanism a supervised one. Although such a learning process may be realized in biomimetic systems of the actomyosin cortex with prescribed external force application in a part of the network, it is not clear how it can happen in a living cell in its native environment. 

Here we shall relax the consideration of having an external supervisor in the learning process and ask if meaningful learning is possible through a process that lacks the a priori knowledge of the desired target strain $\epsilon^*_T$ 
leaving the role of the supervisor obsolete. Cytoskeletal networks are often driven by self-organized actomyosin forces or {\it pulses} in various scenarios such as tissue morphogenesis and cell migration during development~\cite{martin2009,munjal2015,banerjee2017,balaghi2024}. Such actomyosin pulsation can travel spatially~\cite{banerjee2017,munjal2015} leading to asymmetric driving forces by giving rise to fast contractions due to the influx of myosin motors and slow relaxation determined by local turnover timescales. Thus, the forces applied by the supervisor can instead come from such pulses asymmetrically driving the system between the free and clamped states. The clamped state is not set a priori by any external drive but emerges from the local regulation of these pulses. Here we shall consider a single pulse train at the target edge and study the resulting learning dynamics (Fig.~\ref{fig:pulse_lrn}A). 

We consider a case where the source edge is extended and the target edge experiences actomyosin pulsation that contracts the edge periodically. For simplicity, we shall consider the same sawtooth-like temporal dynamics for actomyosin pulsation defined by a fast and slow timescale ($\tau_f$ and $\tau_s$) as we have used before. For pulse trains of constant amplitude $A=A_0$, i.e., the driving force given by Eq.~\ref{eq:driving} becomes $\boldsymbol{f}^e=\lambda(t) \hat{\boldsymbol{r}}$ with $\lambda_\text{max}=A_0$. We find that the target edge learns to contract where the target strain increases with time and is only set by the duration of training as the pulsation has no means of setting a particular target strain value (Fig.~\ref{fig:pulse_lrn}B).

Thus, a meaningful desired target strain value does not arise since there is no coupling between the driving force (the pulses) and the network.
A specific target strain value may emerge in the case where the pulse amplitude is coupled with the state of the network via a quantity such as the target edge length. We find that if the actomyosin forces exerted on the target edge depend on the edge length ($L$), e.g., $A= a_0 L$ such that the driving force becomes $\boldsymbol{f}^e=\lambda(t) \hat{\boldsymbol{r}}$ with $\lambda_\text{max}=a_0 L$ where $a_0 $ is a constant coefficient, the learning dynamics converges to finite target strain values. With the length-dependent driving, a shorter target edge experiences a smaller force from the actomyosin pulsation creating a negative feedback on the driving force. This target edge length-dependent negative feedback on the pulsation amplitude gives rise to specific target strain values in an amplitude-dependent manner (Fig.~\ref{fig:pulse_lrn}C). Although the described learning mechanism does not say how to control the pulsation to train or test when needed, biochemical signaling pathways in the cell may provide such functionalities. It should be noted that a persistent change in the mechanical behavior of the network (i.e., the strain at the target edge) has occurred here via irreversible changes in edge rest length values, indicating that learning has taken place.
Our results suggest that a learning mechanism may emerge from the coupling between the actomyosin pulsation and the cytoskeletal network making it possible to self-organize a physical learning process in a cytoskeletal network of a cell.

\begin{figure}[t]
\begin{center}
    \includegraphics[width=0.95\columnwidth]{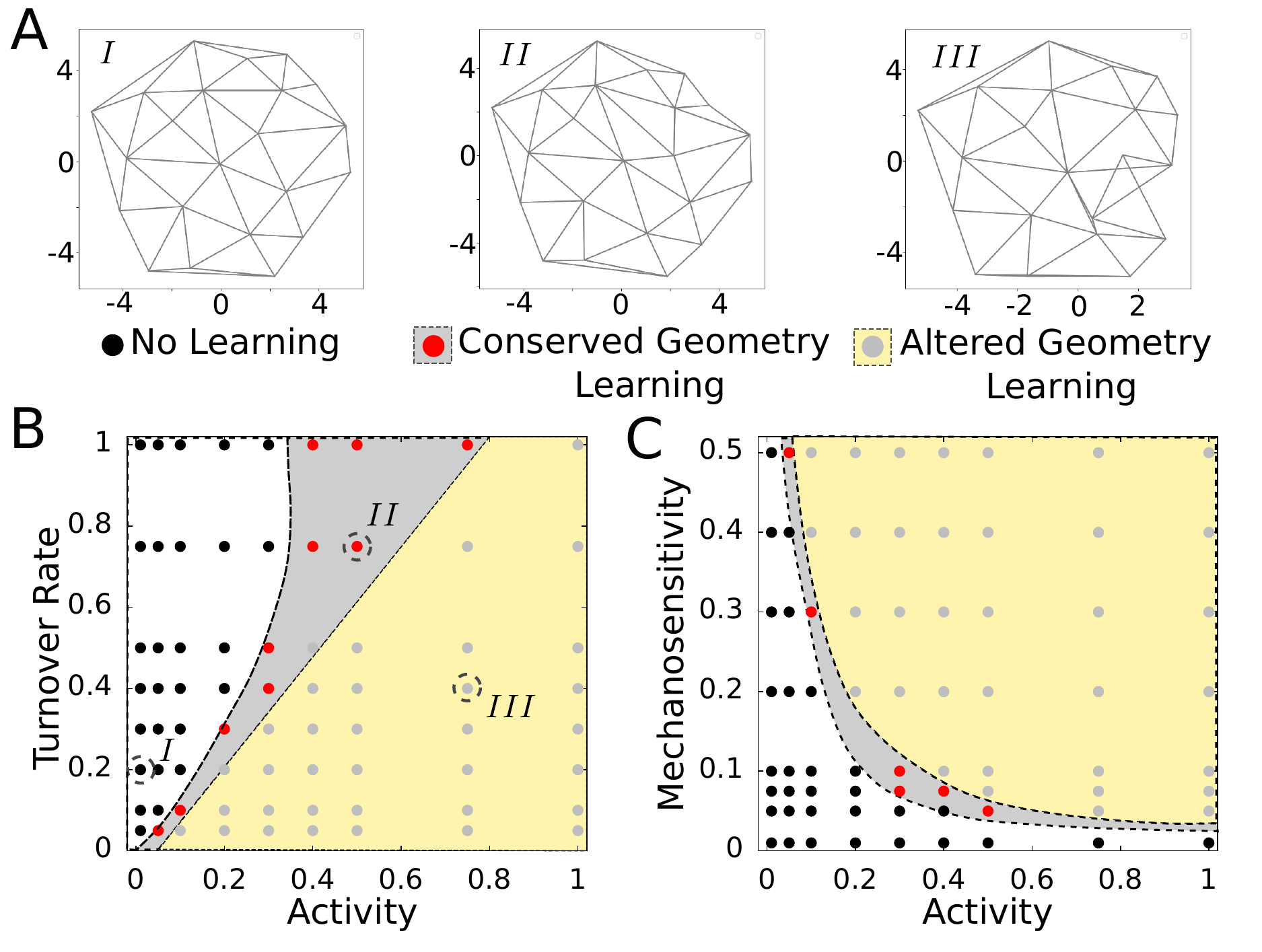}
    \caption{Interplay between motor dynamics, activity and mechanosensitivity enables self-organized learning. (A) Example of trained networks in three different cases corresponding to the points highlighted by broken circles in the phase diagram. (B) Phase diagram showing phases of {\it No learning} (white), {\it Conserved geometry learning} (gray shaded) and {\it altered geometry learning} (yellow shaded) at different values of activity parameter $\xi$ and motor turnover rate $k_u$.
    (C) Phase diagram showing phases of {\it No learning} (white), {\it Conserved geometry learning} (gray shaded) and {\it altered geometry learning} (yellow shaded) at different values of activity parameter $\xi$ and a measure of mechanosensitivity $\beta_1$ (same as $\beta$ in rescaled units). The phases are determined from visualization of trained networks and the boundaries are lines drawn as guides to the eyes. Here $\tau_f/\tau_s=1/4$ and the rescaled parameter values are $\beta=0.1$ (in panel B), $k_u=0.5$ (in panel C), and $\alpha=10$.}
    \label{fig:phase_dia_selforg}
\end{center}
\end{figure}


We further explore the viability of such self-organized learning with negative feedback (Fig.~\ref{fig:pulse_lrn}C) at different regimes of activity, motor dynamics, and mechanosensitivity. We categorize the learning outcomes into three groups: (i) {\it No learning} where the rest lengths do not change at all and no learning occurs, (ii) {\it Conserved geometry learning} where the rest length of the edges locally remodel to give rise to a well-defined target strain with no change in network geometry and (iii) {\it Altered geometry learning} where the rest length changes are large and lead to network geometry changes (Fig~\ref{fig:phase_dia_selforg}A). Note that the network still learns to reach a finite target strain value in the {\it Altered geometry learning} case, but goes through large deformations of some of its edges and changes its geometry. This effect will crucially depend on the nature of the negative feedback and on the network size with smaller networks more prone to larger changes in geometry.
We find that higher activity $\xi$ and lower motor turnover rate $k_u$ (i.e., slower relaxation of motor density) take the network from a non-learning state to conserved geometry and altered geometry learning states (Fig.~\ref{fig:phase_dia_selforg}B). Lower values of mechanosensitivity ($\beta_1$) and activity hinder learning in the network, while increasing values of these two parameters enable learning and take the system from conserved geometry learning at moderate values to altered geometry learning at large values of $\xi$ and $\beta_1$ (Fig.~\ref{fig:phase_dia_selforg}C).

\begin{figure}[t]
\begin{center}
    \includegraphics[width=0.95\columnwidth]{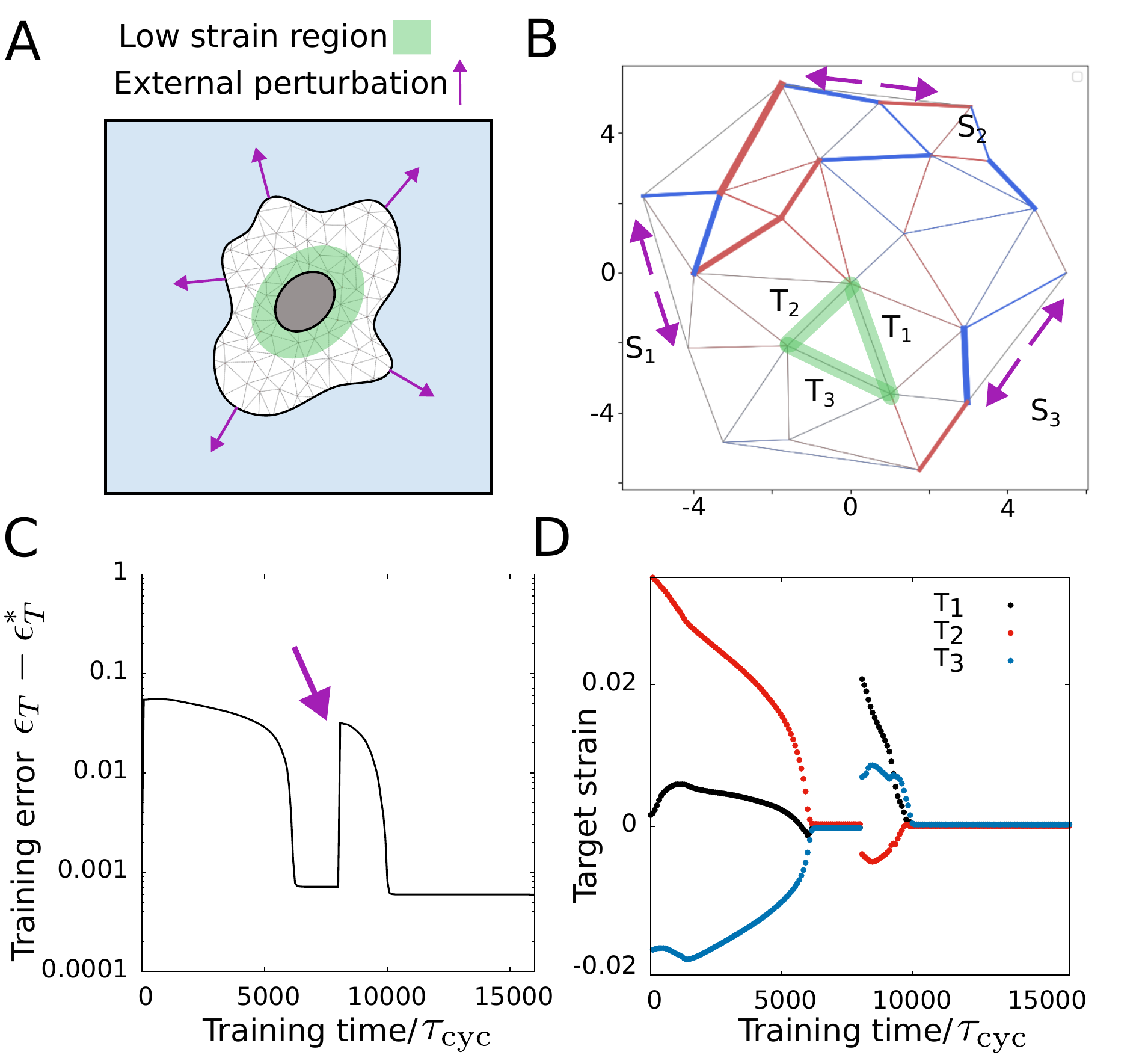}
    \caption{Adaptation via learning. (A) Schematic showing desired low-strain region in the cell in the presence of external mechanical perturbations. (B) The trained network shows the external forces acting on the network (magenta arrows) and the low-strain region in green. (C) Training error reduces obtaining the desired target strain value $\epsilon^*_T=0$ for all target edges.
    (D) Target strain values with training time at three target edges. The $\tau_{\rm cyc}$ and rescaled parameter values are the same as Fig.~\ref{fig:strain_response_l0} except $\alpha=10$.  }
    \label{fig:adaptation}
\end{center}
\end{figure}

\section{Adaptive response from underlying learning process}
The results of the previous sections show how learning can happen in a cytoskeletal network. This capability to learn may aid the cell in various complex tasks that require cellular decision-making, but a clear connection between learning and cellular behavior or response is not present. Previous work hypothesized a relationship between learning in biological systems at the cellular scale and cellular adaptation and homeostasis~\cite{gunawardena2022}. Here, we show that a self-organized physical learning process can lead to an adaptive mechanical response in the network.
The cell nucleus carries the genetic material, the DNA of the cell, and is vital for essential physiological functions of the cell. In a noisy mechanical environment, cellular deformations impart mechanical perturbations on the nucleus via the cytoskeletal network~\cite{haase2016} and may cause DNA damage~\cite{shah2021nuclear}, nuclear rupture and cell death~\cite{denais2016}. To avert such adverse outcomes, the nucleus is known to actively maintain mechanical stability by avoiding large deformations~\cite{rowat2006, wang2018}. The cytoskeletal network surrounding the nucleus is reported to help maintain nuclear stability by reducing its deformability~\cite{wang2018}. Inspired by this observed mechanosensitivity of the nucleus and the need to adapt to mechanical perturbations, we propose an adaptation problem where the cytoskeletal network adapts to lower local strain around the nucleus (i.e., a designated area in the network).

We consider a particular interior region of the network as the perinuclear region and want it to remain in zero or low strain no matter how the network is deformed (Fig.~\ref{fig:adaptation}A). For simplicity, we shall consider three connected edges to constitute the low-strain region and apply deformations at three different boundary edges. We cast this problem into our learning mechanism with the source edges and the target edges, where the deformed boundary edges are equivalent to three source edges ($S_1, S_2, S_3$) and the low-strain region can be considered as three target edges given by $T_1, T_2, T_3$ (Fig.~\ref{fig:adaptation}B). Since we want zero strain, $\epsilon^*_{T_i}=0$ the driving force becomes

\begin{equation}
  \boldsymbol{f}^e_i = -\lambda(t)\boldsymbol{\nabla} (\frac{1}{2}(\epsilon_{T_i})^2 )\, ,
\label{eq:adapt_driving}
\end{equation}
where $i\in 1,2,3$ denotes the target edges. Note that the derivative is in the coordinates of the target node positions and the components of force can be written as $f^e_{ix}=-\lambda(t)\epsilon_{Ti} \cos(\theta)$ and $f^e_{iy}=-\lambda(t)\epsilon_{Ti} \sin(\theta)$ where $\theta$ denotes the orientation of the target edge. It is important to note that since the driving force is completely dependent on the local strain at the target edge, training for this learning problem can be performed via self-organized forces without any external supervisor. Also, since the learning is continuous, the low-strain region can adapt to changing external mechanical perturbations. We find the training error i.e., the total strain in the low-strain region $\sum_i|\epsilon_{T_i}|$ reaches a very low value $\sim0$ as the training progresses (Fig.~\ref{fig:adaptation}C). Upon applying further mechanical perturbation, the strain values in the low-strain region initially increase but gradually return to very small values demonstrating the adaptive response of the network (Fig.~\ref{fig:adaptation}C,D).

\section{Discussion and Conclusion}



Cytoskeletal networks in cells perform a diverse array of functions from providing a basis for force motility, to controlling cell shape, and to providing tracks for intercellular cargo. A large range of  complex, conserved feedback mechanisms enable these networks to perform their task reliably~\cite{haupt2018cells}. Here, we contextualize such phenomenology through the lens of physical learning. In particular, we have shown how a mechanical model inspired by cytoskeletal networks, and endowed with analouges of mechanosensitive proteins and molecular motors, can learn input-output relationships through a form of contrastive learning. While we use simplifying assumptions in building our model, we expect our results to be valid in more complex scenarios such as in the presence of network component turnover and severing.  We also show how this form of learning can lead to or support adaptive mechanical behavior. These results are a proof of principle - in a minimal context - for how bio-molecular mechanisms can support learning processes.

The learning paradigm specified here can, in principle, be applicable to a variety of scenarios. Recent works indicate that the memory of cytoskeletal network configurations controls cell motility and gaits in varying conditions~\cite{kalukula2024, larson2022}. Our learning paradigm might shed light on how mechanosensitivity in addition to external forcing, can allow cytoskeletal networks to encode such information and accurately adapt to changes in the environment. In a broader context, our work can potentially have applications in vertex-like models used to describe various developmental processes. In these scenarios, our work can offer insight to how myosin pulsation can help actuate desired mechanical responses~\cite{noll2017,farhadifar2007influence,alt2017vertex}. 


Our work also offers experimentally testable  predictions. For example, our approach predicts that cells with reduced levels of mechanosensitive proteins should exhibit impaired adaptation to mechanical stimuli. Similarly, we predict that adaptation to mechanical stimuli should increase (decrease) with actomyosin contractility (turnover). 
Recent experimental studies focusing on long-term memory of cellular morphology~\cite{kalukula2024} and the spontaneous local (cellular) strain-dependent process of achieving the desired shape in a drosophila wing disc tissue~\cite{fuhrmann2024} offer the possibility to experimentally test these predictions in the future.

\section{Acknowledgements}

We thank Arvind Murugan for his comments on the manuscript. We thank Rituparno Mandal for providing energy-minimized particle configurations for the initial network creation. DSB thanks Matthew Du and Carlos Floyd for useful discussions. D.S.B and S.V. were supported by the National Institute of General Medical Sciences of the NIH under Award No. R35GM147400. We also acknowledge support from the National Science Foundation through the Physics Frontier Center for Living Systems (PHY-2317138)

\clearpage

\appendix
\renewcommand{\thefigure}{S\arabic{figure}}
\setcounter{figure}{0}

\section{Model Details}
\label{appendix:model}
To understand how physical learning in a cytoskeletal network may occur, we consider the cytoskeletal network as a disordered network of nodes connected by edges.
To create the initial disordered networks, we use an energy-minimized and force-balanced polydisperse particle mixture (with small and big particles in the $1:1$ ratio) with harmonic interaction. The networks are obtained by creating a Voronoi tessellation of the particle positions. The initial network is assumed to be tension-free, i.e., the length of each edge ($L$) is the same as their equilibrium length ($L^0$).
We describe the dynamics of the network by the dynamics of the positions of the nodes $\{\boldsymbol{r}\}$ which governs the length of the edge $L$ and the density of the bound motor $m$ and the bound mechanosensitive protein $n$ at each edge. We define the strain at the edge as $\epsilon=(L-L^0)$. We shall elaborate on the linearization used here by considering the dynamics of one edge in the network.

We consider a force dependent dynamics for the number density of bound mechanosensitive proteins given by
\begin{equation}
  \dot{n} = k_{bn} - k^0_{un} e^{-\beta \dot{\epsilon}} n\, ,
\end{equation}
where $k_{bn}$ and $k^0_{un}$ are the bare binding and unbinding rate and $\beta$ is the coefficient for strain-rate dependent unbinding. 
Mechanosensitive protein dynamics can be very fast with a timescale of a few seconds. Here we consider the strain rate change to be a much slower process than the protein dynamics (i.e., consider constant strain rate when solving for $n$). Now we can linearize the dynamics around a steady state value $n^0$ and zero strain $\epsilon^0=0$ and solve for the variation $\delta n$ as  
\begin{eqnarray}
\delta\dot{n} &=& k_{bn} -k^0_{un} (1-\beta \delta \dot{\epsilon}) (n^0 + \delta n)  \nonumber \\
 &=& \beta n^0 k^0_{un} \delta \dot{\epsilon} - k^0_{un} \delta n  \nonumber \\
 \delta n &=& n^0 \beta \delta \dot{\epsilon} + C_0 e^{-k^0_{un} t} 
\label{eq:limeq}     
\end{eqnarray}
where $n^0=k_{bn}/ k^0_{un}$ is the unstrained steady state mechanosensitive protein density and $C_0$ is the constant of integration.

\begin{figure}[ht!]
\begin{center}
    \includegraphics[width=0.95\columnwidth]{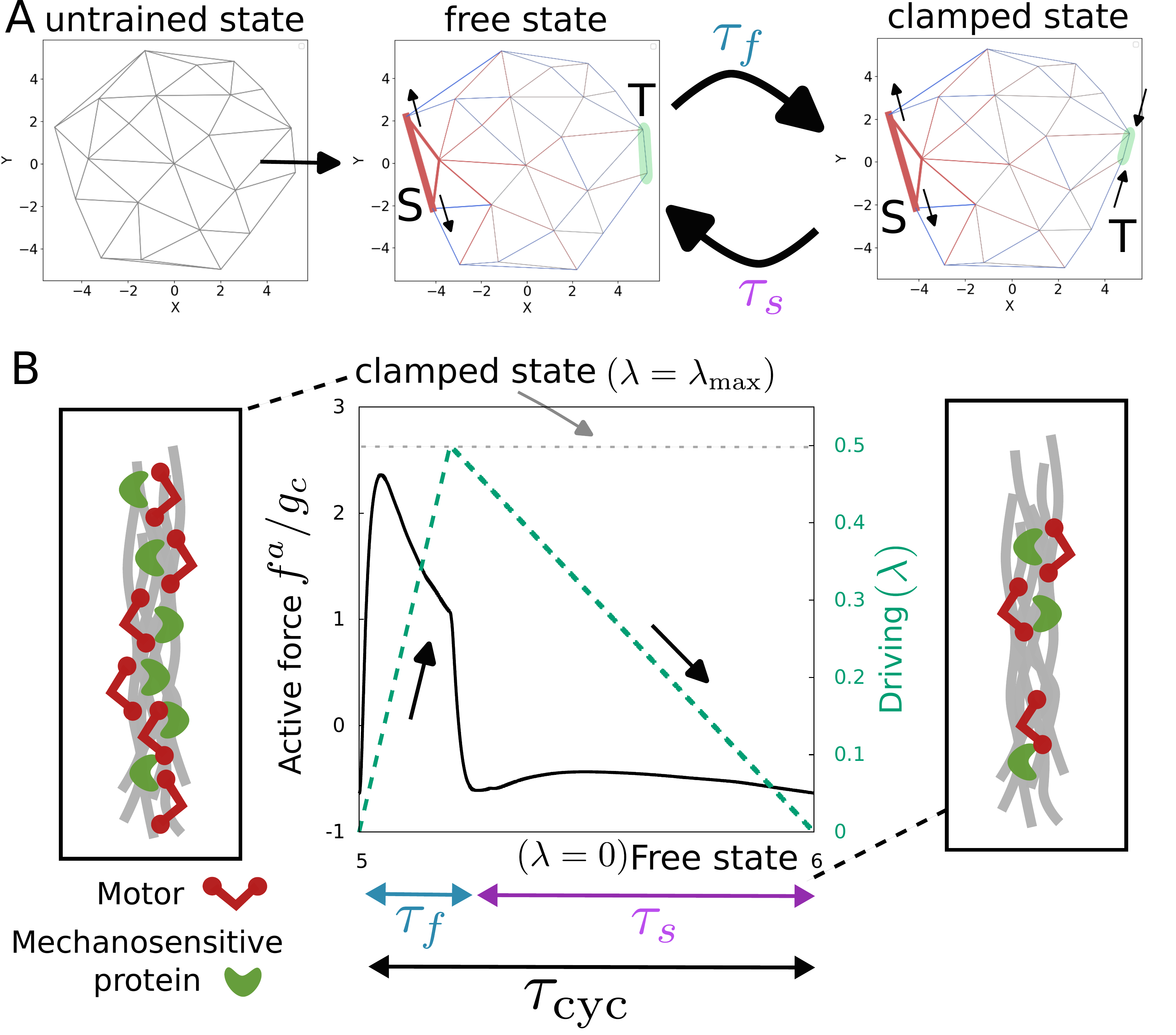}
    \caption{Contrastive learning via driving. (A) The network in free and clamped states. The thickness indicates strain at each edge and the color indicates contraction (blue) or extension(red). (B) Representative active force dynamics (due to change in motor density) in one edge as the target edge is driven from the free to the clamped state over a fast timescale $\tau_f$ and brought back to the free state over a slow timescale $\tau_s$. The motor and mechanosensitive protein density increases when the edge is extended in the clamped state. The figure shows unitless quantities and the rescaled parameter values used in panel B are the same as Fig.~\ref{fig:strain_response_l0}. }
    \label{fig:model_SI}
\end{center}
\end{figure}

The molecular motor binding-unbinding kinetics is known to be mechanosensitive. Here, we consider a mechanosensitive protein-dependent bound motor dynamics given as
\begin{equation}
    \dot{m} = k^0_b + k^1_b n - k_u m 
\end{equation}
where $k^0_b$, $k^1_b$ and $k_u$ are the bare binding and unbinding rates. Now rewriting this above equation in terms of the motor density variation around the steady state value $m^0=(k^0_b + k^1_b n_0)/k_u$ as $m = m^0 + \delta m$ and linearizing the terms we get
\begin{eqnarray}
  \delta\dot{m} & = & k^0_b + k^1_b (n_0 + \delta n) - k_u (m_0 + \delta m) \nonumber \\
  &= & k^1_b \delta n - k_u \delta m \nonumber \\
  &\simeq & k^1_b \beta n_0 \delta\dot{\epsilon} - k_u \delta m \, ,
  \label{eq:deltamyo}
\end{eqnarray}
here we use the steady-state protein variation (as the protein dynamics is fast) to arrive at the above equation.



The dynamics of the learning degree of freedom (i.e., the rest length in this case) of an edge depends on the active force ($f^a$) on that edge which is a function of the motor density variation in the edge and it is given by
\begin{equation}
    \dot{L^0_i} = \alpha g(f^a) \, ,
\end{equation}
where $\alpha$ is the learning rate and $g(x)=x$ for all $|x|\ge g_c$ and $0$ otherwise. The learning in this model is continuous and driven by forcing the system from the free state to the clamped state fast and bringing back slowly over two timescales $\tau_f$ and $\tau_s$ respectively.

Now, the dynamics of the $j^{th}$ node and the adjacent edge between nodes $j$ and $k$ can be described in terms of node position and the variation of motor density as:

\begin{eqnarray}
    \gamma \dot{\boldsymbol{r}}_j &=& \sum_{k}^{nn} -k_{jk}(|\boldsymbol{r}_{j} - \boldsymbol{r}_{k}| - L^0_{jk}) \hat{\boldsymbol{r}}_{jk} + f^a_{jk} \hat{\boldsymbol{r}}_{jk} \nonumber\\
    \delta \dot{m}_{jk} &=&   k^1_b \beta n_0 \delta\dot{\epsilon}_{jk} - k_u \delta m_{jk} \nonumber\\
    \dot{L^0_{jk}} &=& \alpha g(f^a_{jk}) 
\label{eq:bond_dynamics}     
\end{eqnarray}
where $L^0_{jk}$ and $k_{jk}$ are the instantaneous rest length and the stiffness of the edge. The active force in the bond is given by $f^a_{jk}=\xi \delta m_{jk}$. We shall use the disordered network with the above-described dynamics to study physical learning in cytoskeletal networks.

\subsection*{Rescaled dynamical equations}

To derive a set of rescaled dynamical equations, we consider the average distance between the nodes in a network as a length scale $l$ and the duration of a driving cycle $\tau_{\rm cyc}=\tau_f+\tau_a$ as a time scale. We can rewrite the rescaled dynamical equations in terms of unitless parameters as
\begin{eqnarray}
    \dot{\tilde{\boldsymbol{r}}}_j &=& \sum_{k}^{nn} -\tilde{k}_{jk}(|\tilde{\boldsymbol{r}}_{j} - \tilde{\boldsymbol{r}}_{k}| - \tilde{L}^0_{jk}) \hat{\boldsymbol{r}}_{jk} + \tilde{f}^a_{jk} \hat{\boldsymbol{r}}_{jk} \nonumber\\
    \delta \dot{\tilde{m}}_{jk} &=&   \tilde{\beta} \delta\dot{\tilde{\epsilon}}_{jk} - \tilde{k}_u \delta \tilde{m}_{jk} \nonumber\\
    \dot{\tilde{L}}^0_{jk} &=& \tilde{\alpha} g(\tilde{f}^a_{jk}) 
\label{eq:rescaled_dynamics}     
\end{eqnarray}
where $\tilde{\boldsymbol{r}}=\boldsymbol{r}/l$, $\delta\tilde{m}_{jk} = \delta m_{jk}/m_0$, $\tilde{L}^0_{jk}= L^0_{jk}/l$ and $\tilde{f}^a_{jk}=\tilde{\xi} \tilde{m}_{jk}$. The unitless parameters are defined as $\tilde{k}_{jk}=\frac{k \tau_{\rm cyc}}{\gamma}$, $\tilde{\xi}=\frac{\xi m_0 \tau_{\rm cyc}}{\gamma l}$, $\tilde{\beta}=\frac{\beta k^1_b n_0 l}{m_0}$, $\tilde{k}_u=k_u \tau_{\rm cyc}$ and $\tilde{\alpha} = \frac{\alpha \tau_{\rm cyc}^2\xi}{\gamma l^2}$. We shall drop the $\tilde{}$ sign in the discussions to keep notations simple. Here the motor density is rescaled by an equilibrium motor density $m_0$. The average distance in the networks we used is $l\simeq2.5 \, \mu m$. Here, the kernel timescale is determined by the motor turnover timescale $\tau_k=k^{-1}_u$ (see Appendix~\ref{appendix:memory}) and molecular motors like the myosin motor have turnover timescale of a few seconds. For learning via driving at the target, the timescale of driving should be much larger than the timescale of the memory kernel ($\tau_k<\tau_f<\tau_s$) ~\cite{falk2023}. Hence, we consider a cycle duration much larger than motor turnover timescale $\tau_{\rm cyc}=100\, {\rm sec}$ with $\tau_f/\tau_s=1/4$ unless otherwise specified. In the case of classification of environmental signals and self-organized learning we have used $\tau_\text{cyc}=10\, sec$ and $\tau_\text{cyc}=50\, sec$ correspondingly.
We consider an equilibrium motor number density $m_0=10\,{\mu m}^{-1}$ and the factor $k^1_b n_0=1\,{\mu m}^{-1} {\rm sec}^{-1} $. We shall use these described values to derive a set of unitless parameters for our analysis. The threshold value $g_c$ in the activation function $g(x)$ is a hyper-parameter of learning and set at an optimal value in the range of $10^{-6} -10^{-5}$.


\subsection*{Supervised learning}
Here we consider a learning scheme based on the above-described supervised temporal contrastive learning method where the supervisor controls the driving mechanism in which the system is taken from the free state to the clamped state and back according to the desired behaviour (strain at the target edge in this case). This driving force at the target nodes can be described as
\begin{equation}
    \boldsymbol{f} = \lambda(t) \boldsymbol{\nabla} (\frac{\lambda}{2}|\epsilon_T - \epsilon_T^*|^2 )
\end{equation}
where $\epsilon_T$ and $\epsilon^*_T$ are instantaneous and desired strain at the target edge. The function $\lambda(t)$ is a sawtooth function devised according to previously used driving in Falk et al ~\cite{falk2023}. It controls the timescale of driving by incorporating a fast transition (over a time duration $\tau_f$) from free to clamped state and a slow relaxation (over a time duration $\tau_s$) back to the free state. 

\section{Memory kernel}
\label{appendix:memory}
The motor dynamics coupled with the mechanosensitive proteins, possesses the memory of local strain. We can write the motor density variation in the integral form and use the integration by parts to show 
\begin{eqnarray}
    \delta m &=& \beta_1 \int^t_{-\infty} e^{-\frac{(t-t')}{\tau_k} } \delta \dot{\epsilon}(t') dt' \nonumber\\
    &=& \beta_1 \delta \epsilon - \frac{\beta_1}{\tau_k} \int^t_{-\infty} e^{-\frac{(t-t')}{\tau_k} } \delta \epsilon(t') dt' \nonumber \\
    &=& \int^t_{-\infty} \mathcal{K}(t-t') \delta\epsilon(t')  dt'
\label{eq:myo_memory_SI}    
\end{eqnarray}
where the memory kernel is given by 
\begin{equation}
    \mathcal{K}(t-t')=\beta_1 \left( \delta(t-t') - \frac{1}{\tau_k} e^{-\frac{(t-t')}{\tau_k} } \right) \, .
    \label{eq:kernel_SI}
\end{equation}

The consistency between the integral form and the motor dynamics described in Eq.~\ref{eq:bond_dynamics} can be understood if we take a derivative of the above integral form (Eq.~\ref{eq:myo_memory_SI}) and use Leibniz's integral rule
\begin{eqnarray}
 \delta \dot{m} &=& \beta_1 \frac{d}{dt}\left[\int^t_{-\infty} e^{-\frac{(t-t')}{\tau_k} } \delta \dot{\epsilon}(t') dt' \right] \nonumber\\ 
 &=& \beta_1 \frac{d}{dt}\left[\int^t_{-\infty} f(t,t') dt' \right] \nonumber\\
 &=& \beta_1 f(t,t')\biggr\rvert_{t}  \frac{d}{dt}\left( t \right) - \beta_1 f(t,t')\biggr\rvert_{-\infty}  \frac{d}{dt}\left(-\infty \right) \nonumber\\ 
 && + \beta_1 \int^t_{-\infty} \frac{\partial}{\partial{t}}\left(f(t,t') \right) \nonumber\\
 &=& \beta_1 \delta \dot{\epsilon}(t) - \frac{\beta_1}{\tau_k} \int^t_{-\infty} e^{-\frac{(t-t')}{\tau_k} } \delta \dot{\epsilon}(t') dt' \nonumber \\
 &=& \beta_1 \delta \dot{\epsilon}(t) - \frac{1}{\tau_k} \delta m
\end{eqnarray}
which we can now compare with the motor dynamics (Eq.~\ref{eq:bond_dynamics}) to identify the parameters $\tau_k=k_u^{-1}$ and $\beta_1= k^1_b \beta n_0$. In the rescaled parameters, the memory kernel parameters will be slightly different given by $\tilde{\beta}_1 = \tilde{\beta}$ and $\tilde{\tau}_k=\frac{1}{k_u \tau_{\rm cyc}}$.

\section{Learning in nonlinear regime}
\label{appendix:nonlinear}
We have taken a linear approximation in the mechanosensitive protein to derive a motor-dependent implicit memory of local strain. We find that this linear approximation is not a necessary condition for learning. Here we consider higher-order terms in the protein dynamics given by Eq.~\ref{eq:lim} in the maintext to arrive at
\begin{eqnarray}
\delta\dot{n} &\simeq& k_{bn} -k^0_{un} \left(1-\beta \delta \dot{\epsilon} + \frac{\beta^2}{2} \delta \dot{\epsilon}^2 - \frac{\beta^3}{6} \delta \dot{\epsilon}^3\right) (n_0 + \delta n)  \nonumber \\
 &=& \beta n_0 k^0_{un} \phi(\delta \dot{\epsilon}) - k^0_{un} \delta n  
\label{eq:limeq-nonlin}     
\end{eqnarray}
where $\phi(\delta \dot{\epsilon})= \delta \dot{\epsilon} - \frac{\beta}{2} \delta \dot{\epsilon}^2 + \frac{\beta^2}{6} \delta \dot{\epsilon}^3$. We have ignored the contributions of mixed terms (e.g. $\delta n \delta \dot{\epsilon}$) for analytical tractability. Similar to Appendix.~A, we can obtain the solution for protein dynamics given by
\begin{equation}
    \delta n = n_0 \beta \phi(\delta \dot{\epsilon}) + C'_0 e^{-k^0_{un} t} \, ,
\end{equation}
where $C'_0$ is a constant of the integration.

Now, using the steady state solution of protein density variation dynamics given above and the Eq.~\ref{eq:deltamyo}, we arrive at motor density variation dynamics given by
\begin{eqnarray}
  \delta\dot{m} & = & k^1_b \delta n - k_u \delta m \nonumber \\
  &\simeq & k^1_b \beta n_0 \phi(\delta \dot{\epsilon}) - k_u \delta m \nonumber \\
  \label{eq:deltamyo2}
  &=& \beta_1 \delta \dot{\epsilon} - \beta_{2} \delta \dot{\epsilon}^2 + \beta_{3} \delta \dot{\epsilon}^3 - k_u \delta m \, ,
\end{eqnarray}
where $\beta_1=k^1_bn_0\beta$, $\beta_2=k^1_bn_0\beta^2$ and $\beta_3=k^1_bn_0\beta^3$.

\begin{figure}[ht!]
\begin{center}
    \includegraphics[width=0.95\columnwidth]{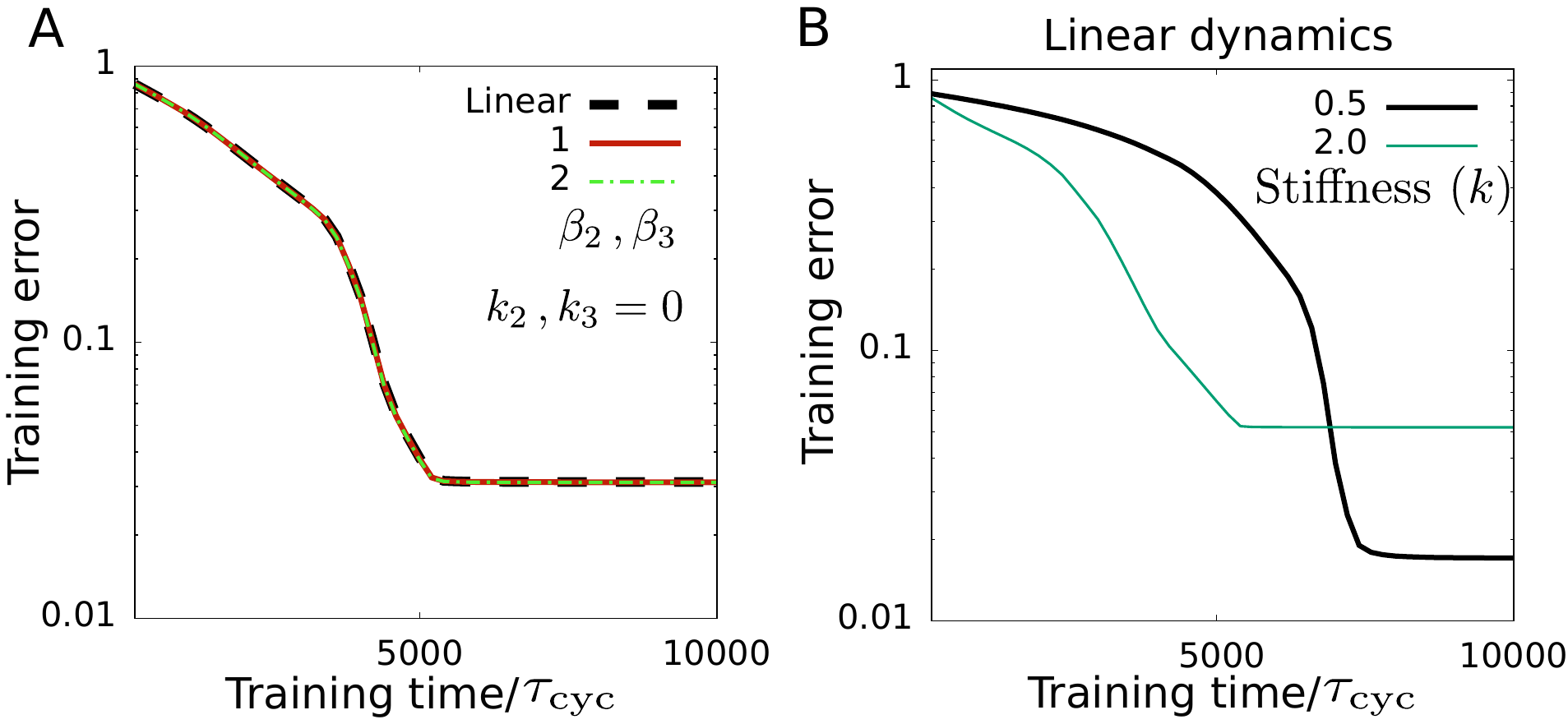}
    \caption{Learning with non-linear mechanosensitive protein dynamics. (A) The temporal evolution of training error does not significantly change in presence of moderate non-linearity. (B) Training error time evolution indicates better learning when the network edges are softer. The parameter values used here are the same as Fig.~\ref{fig:strain_response_l0} except $\lambda_{max}=0.2$ and $\xi=0.2$. }
    \label{fig:nonlinear_SI}
\end{center}
\end{figure}

To understand the implicit memory in the motor dynamics we consider the integral form given by
\begin{eqnarray}
    \delta m &=& \beta_1 \int^t_{-\infty} e^{-\frac{(t-t')}{\tau_k} } \left( \delta \dot{\epsilon} - \frac{\beta}{2} \delta \dot{\epsilon}^2 + \frac{\beta^2}{6} \delta \dot{\epsilon}^3\right) dt' \nonumber\\
    &=& I_1 + I_2 + I_3
\label{eq:myo_memory_SI2}    
\end{eqnarray}
where the integral $I_1$ is same as described in Eq.~\ref{eq:myo_memory_SI} and integrals $I_2$ and $I_3$ can be written as 
\begin{equation}
    I_2=-\frac{\beta}{2}\int_{-\infty}^t \mathcal{K}(t-t') h(t') dt'
\end{equation}
and
\begin{equation}
    I_3=\frac{\beta^2}{6}\int_{-\infty}^t \mathcal{K}(t-t') p(t') dt'
\end{equation}
where $h(t)=\int_{-\infty}^t\delta \dot{\epsilon}^2 dt$ and $p(t)=\int_{-\infty}^t\delta \dot{\epsilon}^3 dt$. We approximate these integrals as 
\begin{eqnarray}
    h(t)&=&\int_{-\infty}^t\delta \dot{\epsilon}^2 dt  \nonumber \\
    &\simeq& \int_{0}^{\epsilon}\delta \dot{\epsilon} d\epsilon \nonumber \\
    &\simeq& \int_{0}^{\epsilon} \frac{\epsilon}{\tau} d\epsilon  \nonumber \\
    &=& \frac{\epsilon^2}{2\tau}
\end{eqnarray}
where we approximate the strain rate $\sim \frac{\epsilon}{\tau}$ using a characteristic timescale $\tau$. We can rewrite the integral form of motor density variation as
\begin{eqnarray}
    \delta m = \int^t_{-\infty} \mathcal{K}(t-t') \left( \epsilon(t') -\frac{\beta}{4\tau} \epsilon^2(t') + \frac{\beta^2}{18\tau^2} \epsilon^3(t') \right) dt' \nonumber \\
\end{eqnarray}
which shows motor density has an implicit memory of a nonlinear function of local strain rather than simply strain. The learning mechanism still works with the same learning rule even though the motor dynamics does not accurately estimate the difference in strain between the free and the clamped state with the above-described form. The training error indicates no significant change in the non-linear regime compared to the linear approximation (Fig.~\ref{fig:nonlinear_SI}A).

Cytoskeletal networks are known to have nonlinear elasitcity. We consider an elastic energy with higher order terms given by
\begin{equation}
    E_{el}=\frac{1}{2} k\epsilon^2 + \frac{1}{3} k_{2} \epsilon^3 + \frac{1}{4} k_{3} \epsilon^4
\end{equation}
where $k_{2}$ and $k_{3}$ are elastic constants corresponding to the higher order terms. The cubic term in the elastic energy originates from activity in the cytoskeletal networks and can take both signs and the quartic term indicates effects like strain-stiffening. Considering nonlinear elasticity, make the contrastive update~\cite{stern2021, falk2023} depend on a nonlinear function of strain 
\begin{eqnarray}
   \dot{L}_0 &=& \left( \frac{\partial{E_{el}}}{\partial L_0} \right|_\text{free} - \left( \frac{\partial{E_{el}}}{\partial L_0} \right|_\text{clamped} \nonumber \\   &=& \left(k\epsilon + k_{2}\epsilon^2 + k_{3} \epsilon^3\right)|_\text{clamped} \nonumber\\ && - \left(k\epsilon + k_{2}\epsilon^2 + k_{3} \epsilon^3\right)|_\text{free} \, . \nonumber \\
\end{eqnarray}
Hence, the implicit memory of a nonlinear function of strain in motor dynamics can now enable contrastive learning for nonlinear elastic interactions.
Together with the non-linearities in mechanosensitive protein dynamics, we find the strength of nonlinearity to affect learning and with increasing strength of non-linearity the training error becomes smaller (Fig.~\ref{fig:strain_response_k}D). The network softening due to the cubic nonlinearity may play a role in affecting learning as a softer network learns better and reaches lower training error (Fig.~\ref{fig:nonlinear_SI}B).

\section*{Supplementary Figures}


\begin{figure}[ht!]
\begin{center}
    \includegraphics[width=\columnwidth]{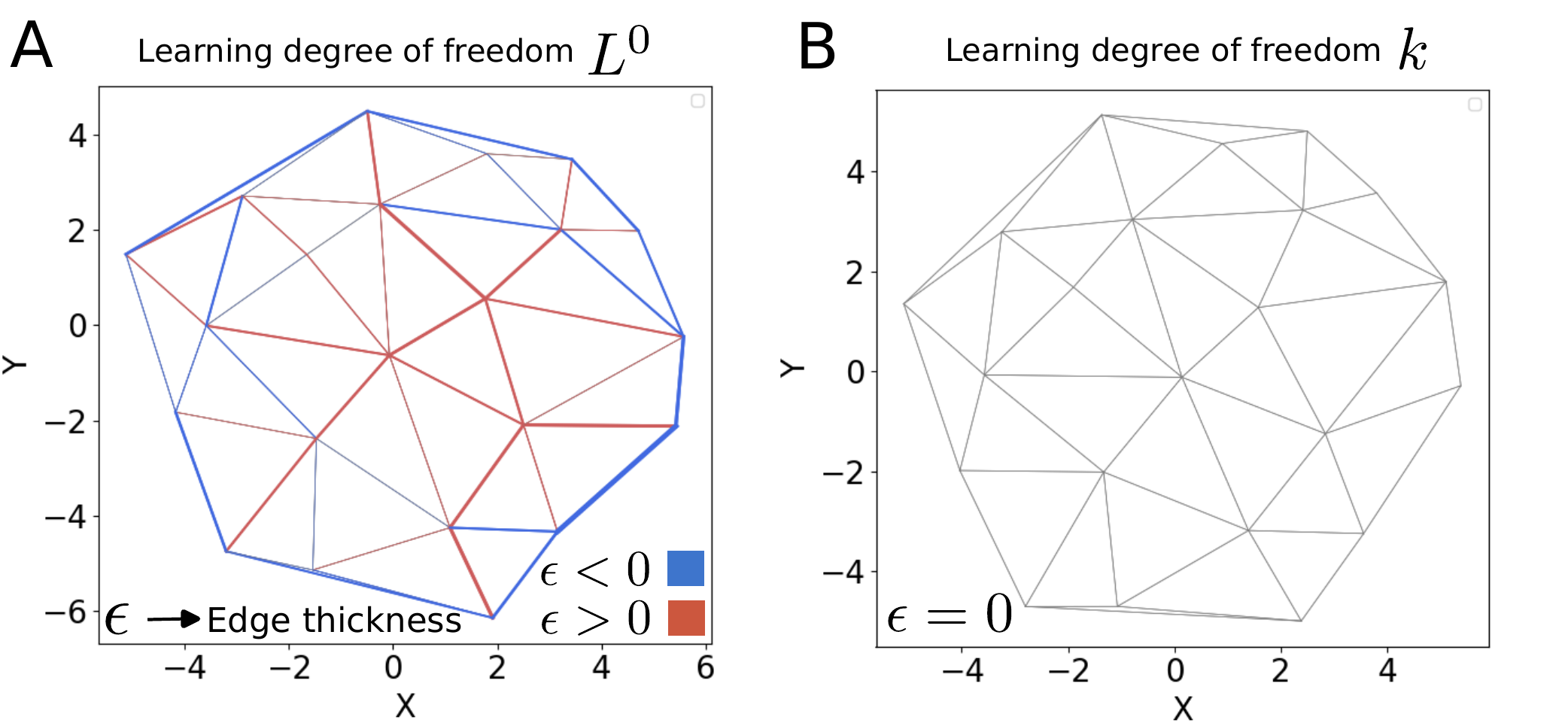}
    \caption{Strain in the trained network. (A) Strain in a trained network with the rest length ($L^0$) as the learning degree of freedom. The edge thickness and colour indicate local strain magnitude and sign. (B) A trained network with the edge stiffness ($k$) as the learning degree of freedom has no strain in the equilibrium state (i.e., without any source strain). The parameters are same described in Fig.~\ref{fig:strain_response_l0} and Fig.~\ref{fig:strain_response_k}. }
    \label{fig:SI_trained_net}
\end{center}
\end{figure}

\begin{figure}[ht!]
\begin{center}
    \includegraphics[width=\columnwidth]{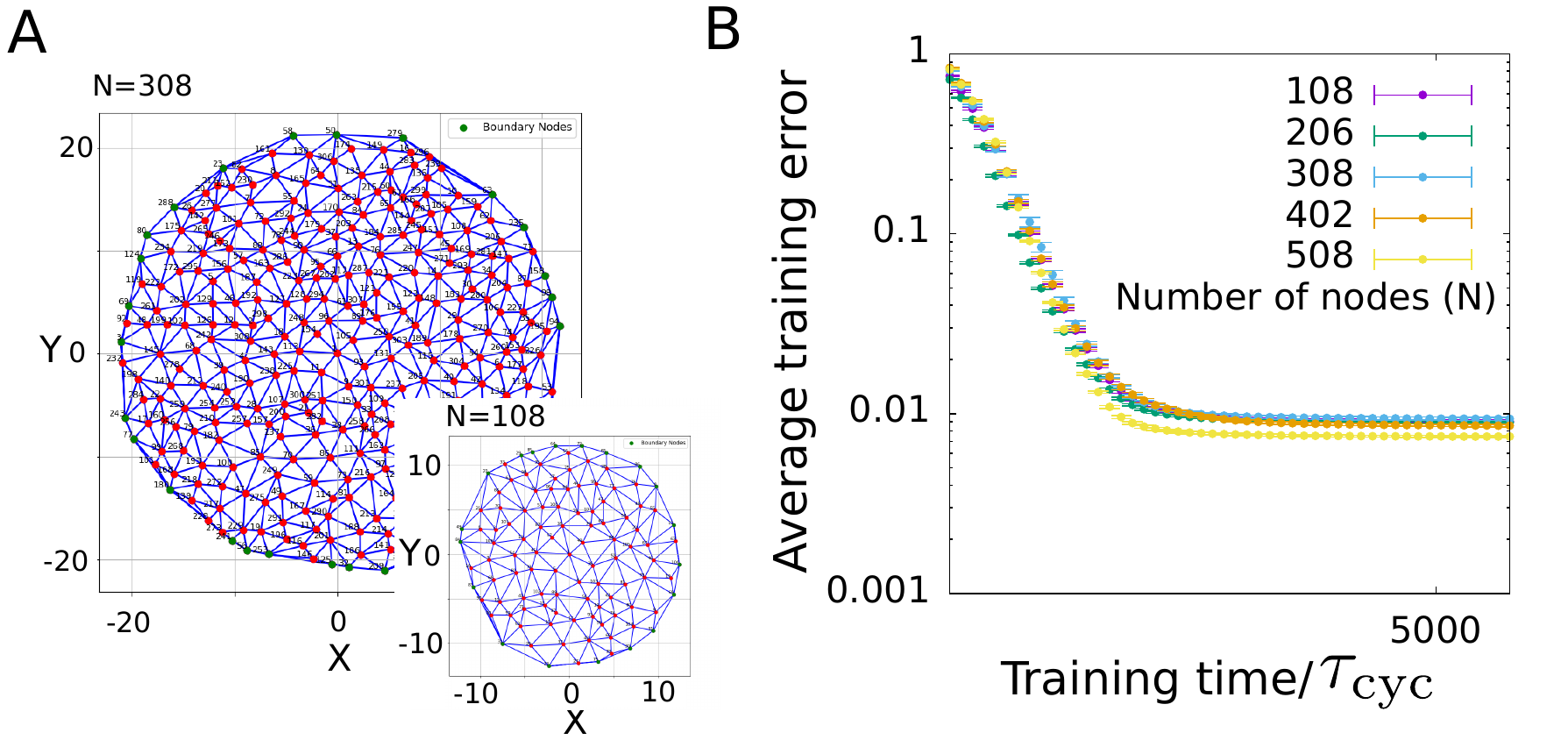}
    \caption{Effect of network size on learning. (A) Networks of different sizes are used for the learning task. One source and one target edge were randomly chosen from the edges on the network periphery. (B) Training error with training time for different sizes shows no significant effect of network size on learning. The parameters used here are the same as Fig.~\ref{fig:strain_response_l0} except $\tau_\text{cyc}=800\, sec$.}
    \label{fig:SI_system_size}
\end{center}
\end{figure}

\vspace{100 mm}

\begin{figure}[ht!]
\begin{center}
    \includegraphics[width=\columnwidth]{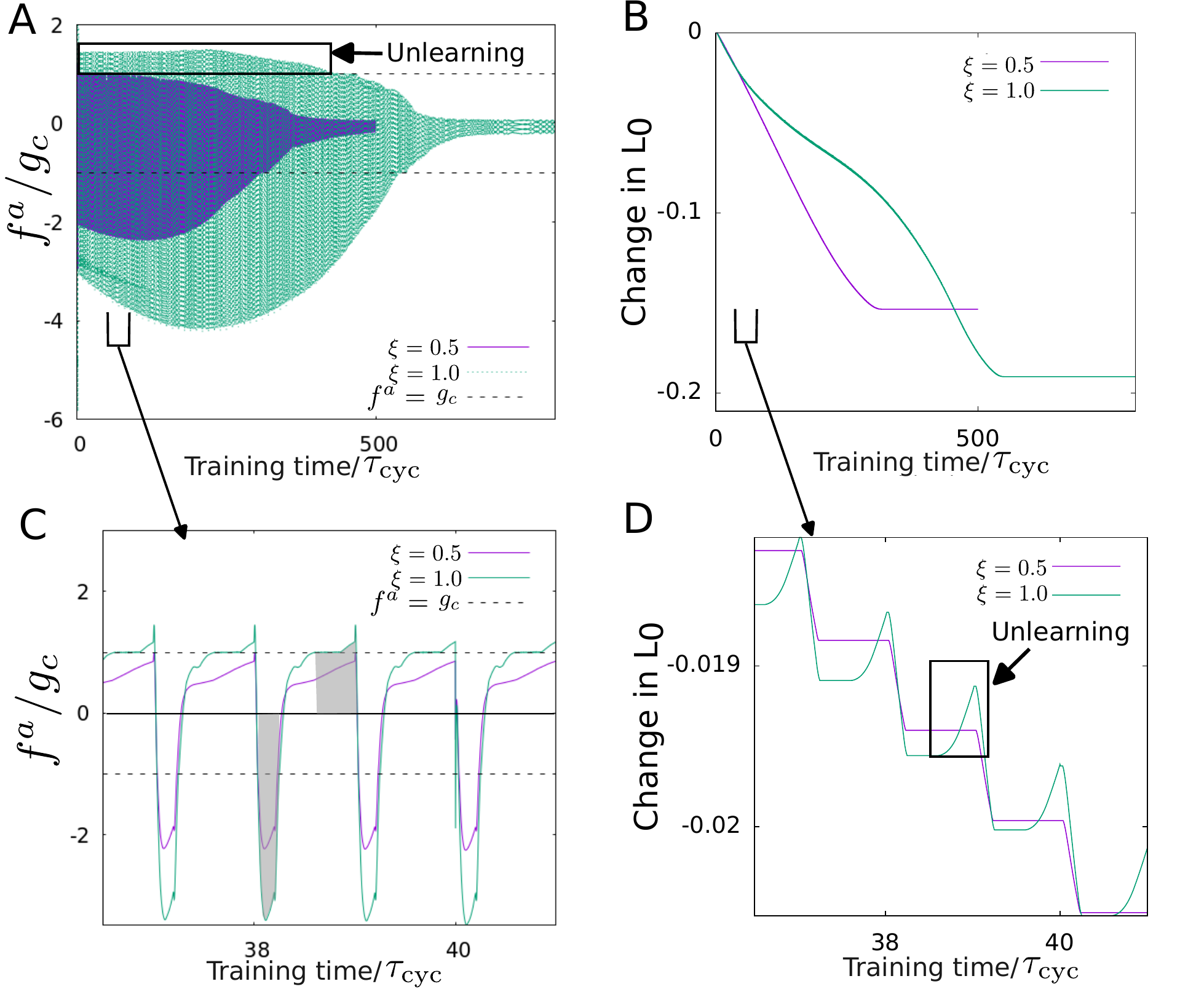}
    \caption{Effect of activity on learning at a single edge. (A) Rescaled active force with time during training in one specific edge of the network. Higher activity shows longer and more learning but also has a significant period of unlearning (highlighted). (B) Higher activity leads to an increased change in rest length ($L^0$) of the edge during training. (C) A zoomed-in version of rescaled active stress vs training time shows the difference in active force at different activity values. The shaded region shows learning (when $f^a<0$) and unlearning (when $f^a>0$). (D) A zoomed-in version of change in rest length during training shows a larger amount of learning in each cycle and a large amount of unlearning at a higher activity value. The parameters used here are the same as Fig.~\ref{fig:strain_response_l0}.}
    \label{fig:SI_unlearning}
\end{center}
\end{figure}

\begin{figure}[ht!]
\begin{center}
    \includegraphics[width=\columnwidth]{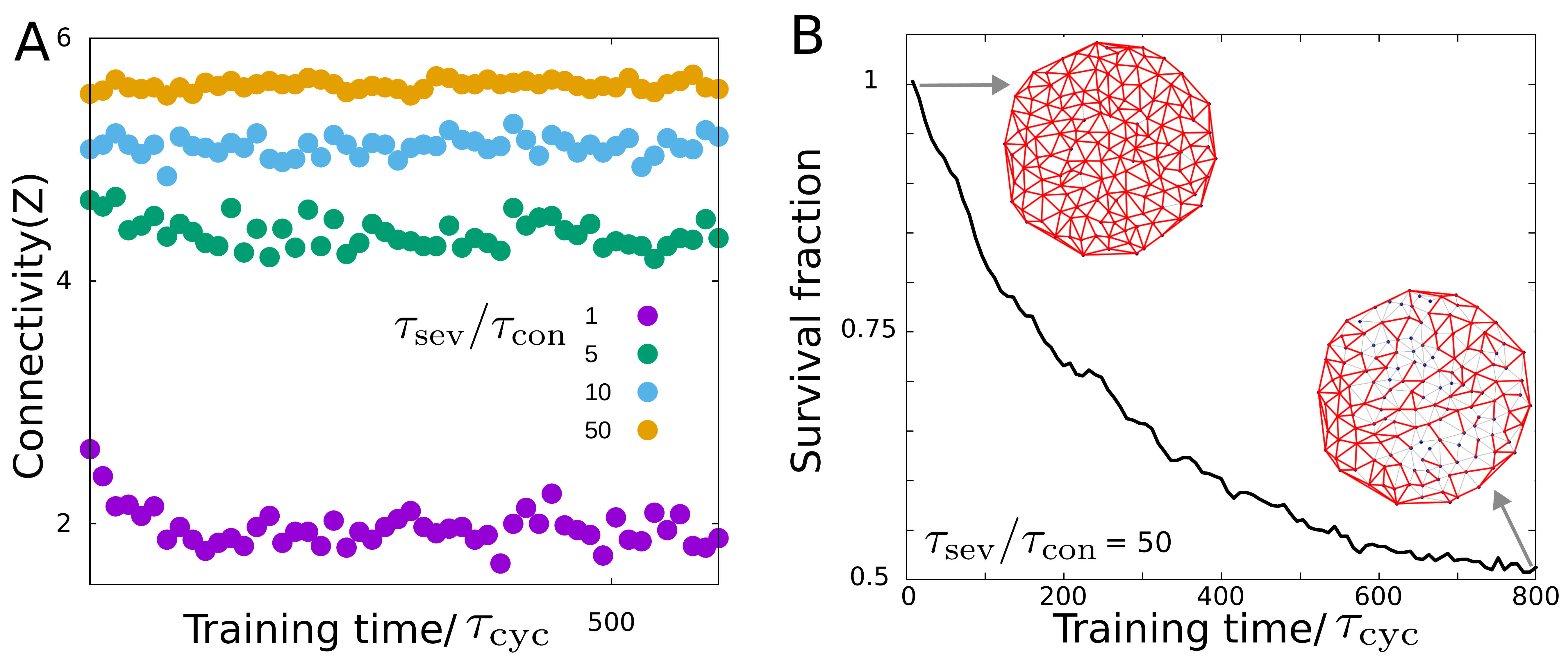}
    \caption{Learning with turnover of network edges. (A) Average network connectivity at different $\tau_\text{sev}/\tau_\text{con}$ ratio during training. (B) Fraction of all edges that have not been severed at any time during the training (survival fraction). Insets show the network at the beginning and ending of training. The edges that have never been severed are indicated in red. Parameter values used here are the same as Fig.~\ref{fig:remodeling}. }
    \label{fig:SI_remodeling}
\end{center}
\end{figure}

\newpage

\bibliographystyle{unsrt}

\clearpage

\end{document}


\title{Supplemental Material: Contractile to extensile transitions and mechanical adaptability enabled by activity in cytoskeletal structures}

\author{Alexandra Lamtyugina*}
\affiliation{Department of Chemistry, University of Chicago, Chicago, IL 60637}

\author{Deb Sankar Banerjee*}
\affiliation{Department of Chemistry, University of Chicago, Chicago, IL 60637}
\affiliation{James Franck Institute, University of Chicago, Chicago, IL 60637}

\author{Yuqing Qiu}
\affiliation{Department of Chemistry, University of Chicago, Chicago, IL 60637}
\affiliation{James Franck Institute, University of Chicago, Chicago, IL 60637}

\author{Suriyanarayanan Vaikuntanathan$^\dagger$}
\affiliation{Department of Chemistry, University of Chicago, Chicago, IL 60637}
\affiliation{James Franck Institute, University of Chicago, Chicago, IL 60637}

\maketitle

\begingroup\renewcommand\thefootnote{$^*$}
\footnotetext{These authors contributed equally to this work.}
\endgroup
\begingroup\renewcommand\thefootnote{$^\dagger$}
\footnotetext{Corresponding author. Email: svaikunt$@$uchicago.edu}
\endgroup

\section{Simulation details}

As a first step towards understanding physical learning in cytoskeletal networks, we shall consider the case of learned mechanical response of the network based on the canonical case of learning allosteric response in mechanical networks~\cite{rocks2017, stern2021}.
We consider the cytoskeletal network as a disordered network of nodes connected by bonds. These bonds can be conceived to be crosslinked bundles of actin filaments that undergo remodeling based on local quantities such as active stress etc. This local remodelling gives rise to malleable material properties in the bonds. We shall consider the bonds to be Hookean springs with stiffness and rest length which can change as the bonds undergo remodelling. This malleability of network properties provides a way to modify the network interaction by local driving. Thus, the rest length and stiffness can be considered to be the learning degrees of freedom (LDF). We describe the network dynamics by the dynamics of edge length $L$, myosin number density $m$ and lim protein number density $n$ (Fig.~\ref{fig:single_bond}). We shall define the strain in the bonds as $\epsilon=(L-L^0)/L^0$.

Let us consider a force dependent dynamics for the number density of bound lim proteins given by
\begin{equation}
  \dot{n} = k_{bn} - k^0_{un} e^{-\beta \dot{\epsilon}} n\, ,
\end{equation}
where $k_{bn}$ and $k^0_{un}$ are the bare binding and unbinding rate and $\beta$ is the coefficient for strain-rate dependent unbinding. 
Now if we consider the strain rate change to be a much slower process than the lim dynamics then we can linearize the lim dynamics around a steady state value $n_0$ and solve for the variation $\delta n$ as  
\begin{eqnarray}
\delta\dot{n} &=& k_{bn} -k^0_{un} (1-\beta \delta \dot{\epsilon}) (n_0 + \delta n)  \nonumber \\
 &=& \beta n_0 k^0_{un} \delta \dot{\epsilon} - k^0_{un} \delta n  \nonumber \\
 \delta n &=& n_0 \beta \delta \dot{\epsilon} + C_0 e^{-k^0_{un} t} 
\label{eq:limeq}     
\end{eqnarray}
where $n_0=k_{bn}/ k^0_{un}$ is the unstrained steady state lim value and $C_0$ is the constant of integration. 

Myosin motor binding-unbinding kinetics is known to be mechano-sensitive. Here we consider a lim-dependent bound myosin dynamics given as
\begin{equation}
    \dot{m} = k^0_b + k^1_b n - k_u m 
\end{equation}
where $k^0_b$, $k^1_b$ and $k_u$ are the bare binding and unbinding rates. Now rewriting this above equation in terms of the myosin variation around the steady state value $m_0=(k^0_b + k^1_b n_0)/k^0_u$ as $m = m_0 + \delta m$ and linearizing the terms we get
\begin{eqnarray}
  \delta\dot{m} & = & k^0_b + k^1_b (n_0 + \delta n) - k^0_u (m_0 + \delta m) \nonumber \\
  &= & k^1_b \delta n - k_u \delta m \nonumber \\
  &\simeq & k^1_b \beta n_0 \delta\dot{\epsilon} - k_u \delta m \, ,
  \label{eq:deltamyo}
\end{eqnarray}
here we consider the steady-state lim dynamics to arrive at the above myosin dynamics.



The dynamics of the learning degree of freedom (i.e., the rest length) depends on the myosin variation in the bonds and it is given by
\begin{equation}
    \dot{L^0_i} = g(\delta m_i) \, ,
\end{equation}
where $g(x)=x$ for all $|x|\ge g_c$ and $0$ otherwise. The learning in this model is continuous and driven by forcing the system from the free state to the clamped state fast and bringing back slowly over two timescales $\tau_f$ and $\tau_s$ respectively. We shall discuss these in detail in the following sections where we discuss particular learning examples.

Now, the dynamics of the $i^{th}$ bond can be described in terms of bond length and the variation of myosin and lim proteins as:
\begin{eqnarray}
    \gamma \dot{L}_i &=& - k_i (L_i-L^0_i) + \sigma^a_i \nonumber\\
    \delta\dot{n}_i &=& \beta n_0 k^0_{un} \delta \dot{\epsilon}_i - k^0_{un} \delta n_i \nonumber\\
    \delta \dot{m}_i &=&   k^1_b \delta n - k_u \delta m \nonumber\\
    \dot{L^0_i} &=& g(\delta m_i) 
\label{eq:bond_dynamics}     
\end{eqnarray}
where $L_i$, $L^0_i$ and $k_i$ are the instantaneous length, rest length and the stiffness of the bond. The active stress/tension in the bond is given by $\sigma^a_i=\xi \delta m_i$. 
Here we have cast the equations in a form such that the integral feedback in the dynamics of myosin becomes clear.
We shall use the disordered network with the above-described dynamics to study physical learning in cytoskeletal networks.

\subsection{Memory kernel}

The myosin dynamics coupled with the network strain possess memory of the deformation. We can write the myosin variation in the integral form and use the integration by parts to show 
\begin{eqnarray}
    \delta m &=& \beta_1 \int^t_{-\infty} e^{-\frac{(t-t')}{\tau_k} } \delta \dot{\epsilon}(t') dt' \nonumber\\
    &=& \beta_1 \delta \epsilon - \frac{\beta_1}{\tau_k} \int^t_{-\infty} e^{-\frac{(t-t')}{\tau_k} } \delta \epsilon(t') dt' \nonumber \\
    &=& \int^t_{-\infty} \mathcal{K}(t-t') \delta\epsilon(t')  dt'
\label{eq:myo_memory}    
\end{eqnarray}
where the memory kernel is given by 
\begin{equation}
    \mathcal{K}(t-t')=\beta_1 \left( \delta(t-t') - \frac{1}{\tau_k} e^{-\frac{(t-t')}{\tau_k} } \right) \, .
    \label{eq:kernel}
\end{equation}

The consistency between the integral form and the myosin dynamics described in Eq.~\ref{eq:bond_dynamics} can be understood if we take a derivative of the above written integral form and use Leibniz's integral rule
\begin{eqnarray}
 \delta \dot{m} &=& \beta_1 \frac{d}{dt}\left[\int^t_{-\infty} e^{-\frac{(t-t')}{\tau_k} } \delta \dot{\epsilon}(t') dt' \right] \nonumber\\ 
 &=& \beta_1 \frac{d}{dt}\left[\int^t_{-\infty} f(t,t') dt' \right] \nonumber\\
 &=& \beta_1 f(t,t')\biggr\rvert_{t}  \frac{d}{dt}\left( t \right) - \beta_1 f(t,t')\biggr\rvert_{-\infty}  \frac{d}{dt}\left(-\infty \right) \nonumber\\ 
 && + \beta_1 \int^t_{-\infty} \frac{\partial}{\partial{t}}\left(f(t,t') \right) \nonumber\\
 &=& \beta_1 \delta \dot{\epsilon}(t) - \frac{\beta_1}{\tau_k} \int^t_{-\infty} e^{-\frac{(t-t')}{\tau_k} } \delta \dot{\epsilon}(t') dt' \nonumber \\
 &=& \beta_1 \delta \dot{\epsilon}(t) - \frac{1}{\tau_k} \delta m
\end{eqnarray}
which we can now compare with the myosin dynamics to identify the parameters $\tau_k=k_u^{-1}$ and $\beta_1= k^1_b \beta n_0$.

Now, we can rewrite the dynamics of learning degree of freedom (LDF) as
\begin{eqnarray}
    \Delta {L}^0_i &=& \int_0^{\tau_s+\tau_f} g \left( \int^t_{-\infty} \delta \epsilon_i(t') \mathcal{K}(t-t') dt' \right) dt \, .
\end{eqnarray}
We shall discuss the learning examples in the following sections using this learning rule.

\subsection{Supervised learning}
Here we consider a learning scheme based on the above-described supervised temporal contrastive learning method where the supervisor controls the driving mechanism in which the system is taken from the free state to the clamped state and back according to the desired behaviour (strain at the target edge in this case). This driving force at the target nodes can be described as
\begin{equation}
    \boldsymbol{f} = \boldsymbol{\nabla} (\frac{\lambda}{2}|\epsilon_T - \epsilon_T^*|^2 ) \times f_\text{saw}(t)
\end{equation}
where $\epsilon_T$ and $\epsilon^*_T$ are instantaneous and desired strain at the target edge. The function $f_\text{saw}(t)$ is a sawtooth function devised according to previously used driving in Falk et al ~\cite{falk2023}. It controls the timescale of driving by incorporating a fast transition (over a time duration $\tau_f$) from free to clamped state and a slow relaxation (over a time duration $\tau_s$) back to clamped state. We shall relax this consideration of supervised learning in a later section.

\section{Effective temperature of filament-motor self-assemblies}

\begin{figure*}[ht!]
\begin{center}
    \includegraphics[width=0.9\linewidth]{si_figures/extension_snapshots/extension_snapshots_v0.4.png}
    \caption{Graphical summary of simulation results, active system ($v_0=0.4$). Gray portions of simulation timelines correspond to regimes with no applied external forces. Red portions of timelines correspond to applied forces. Exact timestamps are shown in gray next to each snapshot. (A) Active asters elongate into bundles for larger $f_\text{ext}$ but eventually break up. (B) Active bundles undergo sustained elongation for smaller values of $f_\text{ext}$. (C) Active bundles undergo unsustained elongation for larger $f_\text{ext}$.}
    \label{si:fig:snapshots_v0.4}
\end{center}
\end{figure*}

\clearpage

\bibliographystyle{unsrt}
\bibliography{references}